\documentclass[onecolumn,prd,aps,superscriptaddress,preprintnumbers,tightenlines,showpacs,nofootinbib,eqsecnum,amsfonts,amsmath,longbibliography,12pt]{revtex4-2}
\usepackage{placeins}
\usepackage{amsmath}
\numberwithin{equation}{section}
\usepackage{mathrsfs}
\usepackage{slashed}
\usepackage{amsfonts}
\usepackage{txfonts}
\usepackage{bbding}
\usepackage[normalem]{ulem}
\usepackage{graphicx}
\usepackage{color}
\usepackage[colorlinks=true, linkcolor=blue, citecolor=cyan]{hyperref}
\newcommand{\bee}{{\bf{e}}}
\newcommand{\be}{\begin{equation}}
\newcommand{\ee}{\end{equation}}
\newcommand{\dd}{\text{d}}
\usepackage{amsmath}
\usepackage[caption=false]{subfig}
\usepackage{bm}
\usepackage{xcolor}
\colorlet{BLUE}{blue}
\newcommand{\beq}{\begin{equation}}
\newcommand{\eeq}{\end{equation}}
\newcommand{\bea}{\begin{eqnarray}}
\newcommand{\eea}{\end{eqnarray}}

\usepackage[stable]{footmisc}

\begin{document}

%\documentclass[12pt,letterpaper]{article}
%\usepackage[letterpaper, margin=1in]{geometry}
%\pdfoutput=1

%\begin{document}

\title{{\it Euclid}: From Galaxies to Gravitational Waves -- Forecasting Stochastic Gravitational Wave Background Anisotropies and Their Cross-Correlation\footnote{This paper is published on behalf of the Euclid Consortium.}}

%%%% Version Wednesday 5th of November 2025 08:40:32 PM UT
\newcommand{\orcid}[1]{}
\author{K.~Z.~Yang}\orcid{0000-0001-8083-4037}
\email{yang5991@umn.edu}
\affiliation{Minnesota Institute for Astrophysics, University of Minnesota, 116 Church St SE, Minneapolis, MN 55455, USA}
\author{G.~Cusin}\orcid{0000-0001-6046-1237}\affiliation{Institut d'Astrophysique de Paris, UMR 7095, CNRS, and Sorbonne Universit\'e, 98 bis boulevard Arago, 75014 Paris, France}\affiliation{Departement of Theoretical Physics, University of Geneva, Switzerland}
\author{V.~Mandic}\orcid{0000-0001-6333-8621}\affiliation{Minnesota Institute for Astrophysics, University of Minnesota, 116 Church St SE, Minneapolis, MN 55455, USA}
\author{C.~Scarlata}\orcid{0000-0002-9136-8876}\thanks{\email{mscarlat@umn.edu}}\affiliation{Minnesota Institute for Astrophysics, University of Minnesota, 116 Church St SE, Minneapolis, MN 55455, USA}
\author{J.~Suresh}\orcid{0000-0003-2389-6666}\affiliation{Universit\'e C\^ote d'Azur, Observatoire de la C\^ote d'Azur, CNRS, Artemis, F-06304 Nice, France}
\author{B.~Altieri}\orcid{0000-0003-3936-0284}\affiliation{ESAC/ESA, Camino Bajo del Castillo, s/n., Urb. Villafranca del Castillo, 28692 Villanueva de la Ca\~nada, Madrid, Spain}
\author{N.~Auricchio}\orcid{0000-0003-4444-8651}\affiliation{INAF-Osservatorio di Astrofisica e Scienza dello Spazio di Bologna, Via Piero Gobetti 93/3, 40129 Bologna, Italy}
\author{C.~Baccigalupi}\orcid{0000-0002-8211-1630}\affiliation{IFPU, Institute for Fundamental Physics of the Universe, via Beirut 2, 34151 Trieste, Italy}\affiliation{INAF-Osservatorio Astronomico di Trieste, Via G. B. Tiepolo 11, 34143 Trieste, Italy}\affiliation{INFN, Sezione di Trieste, Via Valerio 2, 34127 Trieste TS, Italy}\affiliation{SISSA, International School for Advanced Studies, Via Bonomea 265, 34136 Trieste TS, Italy}
\author{M.~Baldi}\orcid{0000-0003-4145-1943}\affiliation{Dipartimento di Fisica e Astronomia, Universit\`a di Bologna, Via Gobetti 93/2, 40129 Bologna, Italy}\affiliation{INAF-Osservatorio di Astrofisica e Scienza dello Spazio di Bologna, Via Piero Gobetti 93/3, 40129 Bologna, Italy}\affiliation{INFN-Sezione di Bologna, Viale Berti Pichat 6/2, 40127 Bologna, Italy}
\author{S.~Bardelli}\orcid{0000-0002-8900-0298}\affiliation{INAF-Osservatorio di Astrofisica e Scienza dello Spazio di Bologna, Via Piero Gobetti 93/3, 40129 Bologna, Italy}
% ...

\begin{abstract}
We estimate the amplitude and spatial anisotropy in the stochastic gravitational wave background (SGWB) energy density due to compact binary coalescence (CBC) events: binary black holes (BBH), binary neutron stars (BNS), and black hole-neutron star (BHNS) mergers. Our starting point is the Flagship Simulation Galaxy Catalogue developed by the Euclid Consortium. For each galaxy in the Catalogue, we use the simulated mass and star-formation to constrain the galaxy's star-formation history, and predict its contribution to the gravitational-wave energy density through CBC mergers. Combining such contributions from all galaxies in the Catalogue results in a prediction for the frequency spectrum and spatial anisotropy of the CBC SGWB. We also compare this prediction to semi-analytical models of SGWB generated by compact binaries. We identify a set of effective parameters that capture the key features of these models, and we apply a Bayesian framework to infer these parameters assuming an ideal scenario of cosmic variance-limited search. This represents the first step toward developing a comprehensive framework that will eventually enable the correlation of SGWB anisotropy and {\it Euclid} galaxy data, potentially allowing us to extract valuable astrophysical information from this new observable.
\end{abstract}

\maketitle
\section{Introduction}
Terrestrial gravitational wave (GW) detectors Advanced LIGO \cite{aLIGO}, Advanced Virgo \cite{avirgo}, and KAGRA \cite{KAGRA:2020agh} have observed over two hundred signals from compact binary mergers \cite{GWTC4} to date, including binary black holes (BBH), binary neutron stars (BNS), and neutron star-black hole (NSBH) systems. These observations have enabled measurements of rates and distributions of binary systems \cite{GWTC4pop}, tests of General Relativity \cite{O3TGR}, independent measurements of the Hubble constant \cite{GWTC4H0}, and examinations of the neutron star equation of state \cite{GW170817_EOS}, among others. 

One of the most compelling targets for the future observing runs of terrestrial GW detectors is the stochastic gravitational-wave background (SGWB), the superposition of many uncorrelated signals \cite{maggiore,regimbau_review}. Because SGWB inherits anisotropies from the large-scale distribution of its sources, cross-correlating SGWB anisotropies with maps of galaxy number counts is crucial for extracting astrophysical and cosmological information and for mitigating Poisson shot noise \cite{Jenkins:2019uzp,Jenkins:2019nks, Alonso:2020mva, Cusin:2019jpv, Yang:2020usq, Yang:2023eqi, Cusin:2018rsq, Capurri:2021zli, Alonso:2024knf,Sah:2023bgr,Mukherjee:2019oma, Ferraiuolo2025} (see also \cite{Cusin:2025xle, Pitrou:2024scp} for recent Pulsar Timing Arrays experiments). In this work, we highlight the complementarity between SGWB and galaxy maps by leveraging the {\it Euclid} Flagship Simulation Galaxy Catalogue \cite{Euclid:2024few}, to predict both the SGWB angular power spectrum and its cross-correlation with the galaxy distribution, and by identifying the key astrophysical parameters that these joint measurements can constrain. The full impact of {\it Euclid} observational systematics will be addressed in a future work. 

The {\it Euclid} Wide Survey (EWS) will deliver quasi--full-sky coverage ($\sim 14\,000\,\mathrm{deg}^2$ of low-extinction extragalactic sky) with uniform, space-based optical+NIR imaging and slitless NIR spectroscopy. This combination provides exactly the ingredients required to cross-correlate SGWB anisotropies with galaxy maps: $i)$ Very wide angular coverage, enabling robust recovery of the lowest multipoles well matching the expected resolution of the SGWB; $ii)$ high galaxy number density with tomographic photo-$z$ bins for shot-noise suppression and redshift localization of the cross-correlation signal; and $iii)$ stable, homogeneous systematics control (PSF, depth, completeness, stellar contamination, and dust extinction) across a contiguous footprint that overlaps the terrestrial GW network.
Taken together, these properties make the EWS not merely suitable but eventually the ideal, and possibly unique, dataset for a mature SGWB-galaxy cross-correlation measurement, combining near--all-sky angular reach with the depth, number density, and calibration fidelity needed to turn a detection into precise astrophysical and cosmological inference.

The SGWB is expected to include contributions from a wide array of early-Universe processes, primordial tensor fluctuations \cite{grishchuk,barkana,starob,turner}, inflation with gauge-field backreaction \cite{peloso_parviol,seto}, post-inflationary parametric resonance \cite{eastherlim}, and cosmic (super)strings \cite{caldwellallen,DV1,DV2,cosmstrpaper}, as well as from numerous astrophysical sources such as compact binary coalescences (BBH, BNS, NSBH) \cite{regfrei,zhu_cbc,marassi_cbc}, rotating neutron stars \cite{cutler,bonazzola,marassi_magnetar}, and supernovae \cite{regimbau_review,SNe}. While cosmological components are typically stationary and continuous over observation time, the astrophysical background in the frequency band of ground-based detectors is “popcorn-like,” reflecting the discrete and short-lived nature of mergers in band. Consequently, the SGWB angular power spectrum from compact binaries contains a prominent Poisson shot-noise term in addition to clustering \cite{Cusin:2019jpv, Jenkins:2019uzp, Jenkins:2019nks, Alonso:2020mva}. Cross-correlating a shot-noise–dominated SGWB map with a dense galaxy map has been shown to alleviate shot-noise limitations, enabling sharper inference from Earth-based GW detectors \cite{Jenkins:2019uzp,Jenkins:2019nks, Alonso:2020mva, Cusin:2019jpv, Yang:2020usq, Yang:2023eqi, Cusin:2018rsq, Capurri:2021zli, Alonso:2024knf}.

To model SGWB anisotropies induced by large-scale distribution of astrophysical sources, one may adopt semi-analytic prescriptions for the galaxy distribution (e.g., \cite{Cusin:2018rsq, Cusin:2019jhg}) or populate large mock galaxy catalogues using recipes for compact-binary formation and merger rates \cite{Jenkins:2018kxc}. In the latter approach, the growth of dark matter structure is first simulated and the history of the hierarchical mass assembly is then recorded by means of merger trees of haloes and subhaloes within snapshots stored at different time steps.

In this paper, we use the {\textit{Euclid}} Flagship Simulation Galaxy Catalogue (version 2.1.10) \cite{Euclid:2024few} to sample the galaxy distribution. The Catalogue includes 4.8 billion galaxies across one octant of the sky and at redshifts up to 3. We implement a star-formation history (SFH) model $\psi(z)$ for each galaxy, driven by its current star-formation rate and total stellar mass specified in the Catalogue. We combine the SFH model with a distribution of time delay $p(t_{\text{d}})$ between binary formation and merger and with the distribution of metallicity $p(\mathcal{Z})$ to obtain the merger rate of compact binaries hosted by the galaxies. We use the estimates of the mass distribution for compact binaries, provided by Advanced LIGO and Advanced Virgo observations through GWTC-3 \cite{KAGRA:2021duu}, and follow the method of \cite{Jenkins:2018kxc} to compute the SGWB energy density due to the entire population of merging compact binaries. We predict both the angular power spectrum of the resulting SGWB anisotropy and its cross-correlation with the galaxy distribution. These angular spectra, therefore, encode the relevant astrophysical information defined by the SFH $\psi$, delayed time distribution $p(t_{\text{d}})$, and metallicity distribution $p(\mathcal{Z})$. In other words, the measurements of the SGWB angular power spectrum and the SGWB-galaxy correlation angular power spectrum can be used to constrain the combination of $\psi$, $p(t_{\text{d}})$, and $p(\mathcal{Z})$. Combining these measurements with other observations (e.g. SFH $\psi$ measurements by {\it Euclid}) can then further constrain the remaining quantities: $p(t_{\text{d}})$ and $p(\mathcal{Z})$. 

We also compare the angular spectra computed using the {\it Euclid} Flagship Simulation Galaxy Catalogue with the spectra predicted by semi-analytic astrophysical models, such as the ones explored in \cite{Cusin:2019jpv, Cusin:2019jhg}. We identify a set of effective parameters that capture the main features of the semi-analytic models, and we apply a Bayesian formalism to infer their values based on the comparisons of the models with the simulation based on the Catalogue. This approach enables us to identify the range of (effective) astrophysical parameters consistent with the predictions derived from the Catalogue. When applied to real data, this methodology will help determine which class of semi-analytic models and parameters best describe observations.
%This result can then be used to further develop the semianalytic models and constrain the structure formation implemented in the models. 
%Parameterizing with a Gaussian function the astrophysical kernel describing the local process of GW emission at galactic scales, we find the 95\% upper limit on kernel amplitude to be \giulia{to be completed}

This paper is structured as follows. In Sect. \ref{Sec:TheoryModel} we present the astrophysical model of angular power spectra of SGWB fluctuations, galaxy over-density and their cross-correlations. In Sect. \ref{sec:Eqs} we review the approach of \cite{Jenkins:2018uac} of populating a galaxy catalogue with compact binary mergers and predicting the corresponding SGWB spectrum. We present the properties of the {\textit{Euclid}} Flagship Simulation Galaxy Catalogue in Sect. \ref{sec:Euclid} and the SGWB angular power spectra predicted by it in Sect. \ref{sec:PredCBC}. We conclude in Sect. \ref{sec:discussion}.

\section{Modelling Anisotropies and Cross-correlations}
\label{Sec:TheoryModel}

The observed GW energy density parameter, $\Omega_{\text{GW}}$ is defined as the background energy density $\rho_{\text{GW}}$ per units of logarithmic frequency $f$ and solid angle $\bf{e}=(\theta,\phi)$, normalized by the critical density of the Universe today $\rho_{\rm c}$ and the speed of light $c$. It can be divided into an isotropic background contribution $\bar{\Omega}_{\text{GW}}$ and a contribution from anisotropic perturbations $\delta \Omega_{\text{GW}}$ \cite{Cusin:2017fwz, Cusin:2017mjm, Cusin:2019jpv, Pitrou:2019rjz, Cusin:2018avf} as
\begin{align}
	\Omega_{\text{GW}}({\bf{e}},f)=\frac{f}{\rho_{\rm c}\,c}\,\frac{\dd^3 \rho_{\text{GW}}}{\dd^2 {\bf{e}}\,\dd f}({\bf{e}},f)=\frac{\bar{\Omega}_{\text{GW}}(f)}{4\pi}+\delta \Omega_{\text{GW}}({\bf{e}},f)\,,
\end{align}
where the isotropic background spectrum can be written as the integral over conformal distance $r$
\begin{equation}\label{BackandPert}
	\bar{\Omega}_{\text{GW}}(f) = \int \dd r \; \partial_r\, \bar{\Omega}_{\text{GW}}(f,r) = \int \dd r\; \frac{f}{\rho_{\rm c}\,c}\,\mathcal{A}(f,r)\,.
\end{equation}
The integral in Eq. (\ref{BackandPert}) runs over an interval of time $[r_*, r_{\text{o}}]$ where $r_{\text{o}}$ is the conformal distance, and $r_*$ is the maximal distance, which in our case, is chosen to be the distance at a maximum redshift $z=7$. The function $\mathcal{A}(f,r)$ is an astrophysical kernel that contains information on the local production of GWs at galaxy scales. Schematically this kernel can be parameterized as \cite{Alonso:2020mva} 
\begin{equation}
	\mathcal{A}(f,r)=\frac{a^4}{4\pi}\,\int \dd \mathcal{L}_{\text{GW}}\;\bar{n}_{\rm{G}}(\mathcal{L}_{\text{GW}}, r)\, \mathcal{L}_{\text{GW}}\,,
\end{equation}
where $a$ is the scale factor of the Universe and $\bar{n}_{\rm{G}}$ is the average physical number density of galaxies at distance $r$ with gravitational wave luminosity $\mathcal{L}_{\text{GW}}$. Different astrophysical models give quite different predictions for this kernel, see e.g. \cite{Cusin:2019jpv} for an explorative approach. For the SGWB due to mergers of compact objects such as BBH and BNS, the low-frequency band ($f \lesssim 100$ Hz) is dominated by the inspiral phase contributions and follows a simple power law $\Omega_{\rm GW} \sim f^{2/3}$. The redshift dependence of the astrophysical kernel can be parameterized as a Gaussian distribution with parameters $\theta$, which will be discussed in Section \ref{subsec:cross-corr}.

Since $\delta\Omega_{\text{GW}}$ is a stochastic quantity, it can correlate with other cosmological stochastic observables. An interesting observable to look at is the cross-correlation of the SGWB with the distribution of galaxies, for example, with the galaxy number counts $\Delta$ defined as the overdensity of the number of galaxies per unit of redshift and solid angle
\begin{equation}\label{eq:Ncounts}
    \Delta(\bee, z)\equiv \frac{N(z, \bee)-\bar{N}(z)}{\bar{N}(z)}\,,
\end{equation}
where $\bar{N}(z)$ is the average galaxy number density at redshift $z$.

If astrophysical GW sources are located in galaxies, we would expect the SGWB and the galaxy distribution to have a high correlation level. 
The angular power spectrum of the GW and galaxy counts cross-correlations is defined as
\begin{equation}\label{eq:modelCldef}
    C^{\text{cross}}_{\ell}(f; \theta)\equiv \frac{1}{2\ell+1}\,\sum_{m=-\ell}^{\ell}\,\langle a_{\ell m}(f; \theta)\,b^*_{\ell m}\rangle\,,
\end{equation}
where the brackets denote an ensemble average while $a_{\ell m}(f)$ and $b_{\ell m}$ are the coefficients of the spherical harmonics decomposition of the SGWB energy density and galaxy number counts, respectively. Explicitly 
\begin{align}\label{Eq:Sec2Sph}
    \delta \Omega_{\text{GW}}({\bf{e}},f; \theta) &= \sum_{\ell=0}^{\infty} \sum_{m=-\ell}^{\ell} a_{\ell m}(f; \theta) \, Y_{\ell m}({\bf{e}})\,,\\
    \Delta({\bf{e}}) &= \sum_{\ell=0}^{\infty} \sum_{m=-\ell}^{\ell} b_{\ell m} \, Y_{\ell m}({\bf{e}})\,.
\end{align}
It can be shown that the angular power spectra of the cross-correlation are given by \cite{Cusin:2017fwz} 
\begin{equation}\label{eq:modelCl}
	C_{\ell}^{\text{cross}}(f; \theta)= \frac{2}{\pi}\int \dd k\; k^2 \, \delta \Omega^{*}_{\text{GW}\,,{\ell}}(k,f; \theta) \, \Delta_{\ell}(k)\,,
\end{equation}
where $k$ is the wave number. 
Keeping only the leading-order contribution to the anisotropy given by clustering (neglecting line-of-sight effects), we have
\begin{align}
    \delta \Omega_{\text{GW}\,,\ell}(k,f; \theta)&=\frac{f}{4\pi \rho_{\rm c}\,c} \int \dd r \, \mathcal{A}(r,f; \theta) \, \big[b(r)\,\delta_{{\rm m},k}(r)j_{\ell}(k r) \big]\,,\\
    \Delta_{\ell}(k)&=\int \dd r \, W(r)\, \big[b(r)\,\delta_{{\rm m},k}(r)j_{\ell}(k r) \big]\,, 
\end{align}
where $j_{\ell}$ are spherical Bessel functions, while $\delta_{{\rm m}}$ is the dark-matter over-density, related to galaxy overdensity via the bias factor $b$, which we assume to be scale-independent with redshift evolution given by $b(z)=b_0 \sqrt{1+z}$ \cite{WiggleZ:2013kor, Rassat:2008ja} and $b_0=1.5$ \cite{Cusin:2019jpv}, where this value has been set considering the measurement from the SDSS survey at $z=0.6$. %thus the correlation function in \cite{Cusin:2019jpv} matches that of SDSS in \cite{SDSSIIIclustering} at z=0.6. 
$W(r)$ is a window function normalized to one, which we choose to be flat, so it is equivalent to a cut of the observational redshift limit, in our case at $z=3$.

\section{Method Review: Predicting CBCs from a Catalogue}\label{sec:Eqs}

We aim now to compute the background energy density due to GW emissions by compact binary coalescences (CBCs) scattered across galaxies in a galaxy catalogue. Following the method presented in section IV of \cite{Jenkins:2018uac}, we calculate the energy density by summing up contributions from each galaxy, using 
\begin{equation}
    \Omega_{\text{GW}}({\bf{e}},f) = \sum_k \varw_k (f) \,\delta_{\rm D}^{(2)}({\bf{e}}, {\bf{e}}_k)\,,
\end{equation}
where $\delta_{\rm D}^{(2)}({\bf{e}}, {\bf{e}}_k)=\delta_{\rm D}(\cos\theta-\cos\theta_k)\,\delta_{\rm D}(\phi-\phi_k)$ is a 2-dimensional Dirac delta for the direction in the sky. The contribution from each galaxy is given by 
\begin{equation}\label{eq:OGW}
    \varw_k(f)\equiv \sum_i \frac{\pi H_0}{3}\,(t_H\,f_{\text{o}})^3\,\frac{1+z_k}{r^2(z_k)}\,(1+{\bf{e}}_k\cdot \bm{\varv}_{\text{o}}) \int \dd \zeta_{\text{b}} \; R_i(z_k,\mathcal{Z}_k,\zeta_{\text{b}})\,\mathcal{S}_i(f_{\text{s},k},\zeta_{\text{b}})\,.
\end{equation}
%
%\begin{equation}\label{eq:OGW}
%    \Omega_{\text{GW}} (f_{\text{o}},\hat{\bf{e}}_{\text{o}})
%    =\sum_k \sum_i \frac{\pi H_0}{3}\,(t_Hf_{\text{o}})^3 \,\frac{1+z_k}{r^2(z_k)}\,(1+\hat{\bf{e}}_k\cdot \bm{\varv}_{\text{o}}) \int \dd \zeta_{\text{b}} \; R_i(z_k,\mathcal{Z}_k,\zeta_{\text{b}})\,\mathcal{S}_i(f_{\text{s},k},\zeta_{\text{b}})\,\delta^{(2)}({\hat{\bf{e}}}_{\text{o}},{\hat{\bf{e}}}_k).
%\end{equation}
%
In this equation, $R_i$ denotes the merger rate and $\mathcal{S}_i$ the source function, whose physical interpretation will be discussed below, and $t_H\equiv 1/H_0$ stands for the Hubble time. The index $i\in\{$BNS, BBH, BHNS$\}$ runs over the different types of binary mergers: binary neutron star systems, binary black hole systems, and black hole-neutron star systems. These binaries are described by parameters $\zeta_{\text{b}}$, which we will assume to include the component masses $m_{1,2}$ and their spins $\chi_{1,2}$. Furthermore, $k\in \{1,2,...,\mathcal{N}\}$ indexes a sum over the galaxies in the catalogue, with each galaxy $k$ described by its redshift $z_k$, comoving distance $r(z_k)$, sky location $\hat{\bf{e}}_k$, and log-normalized metallicity $\mathcal{Z}_k$ defined relative to the Solar metallicity $Z_\odot$
\begin{equation}\label{eq:logZ}
    \mathcal{Z}\equiv \log_{10}\frac{Z}{Z_\odot}\,.
\end{equation}
%\giulia{I think that what you write here below is actually an exact relation as long as $z_k$ is the observed redshift (but is it here?), see how I would rephrase}\sout{We approximate the source-frame GW frequency in Eq. (\ref{eq:OGW}) as $f_{\text{s},k} \approx f_{\text{o}}\,(1+z_k)$, ignoring the effects of the galaxy peculiar velocity which is expected to be negligible at high redshifts. We also assume that the observer is stationary, setting $\bm{\varv}_{\text{o}} \approx 0$.}
In Eq. (\ref{eq:OGW}), the source-frame GW frequency in galaxy $k$ is related to the observed one via a redshift factor $f_{\text{s},k}= f_{\text{o}}\,(1+z_k)$, where the subscript ``s" represents ``source-frame", and the observed redshift is related to the background one as $z_k=\bar{z}_x+\delta z$, where the perturbation $\delta z$ accounts for all relativistic effects, and in particular for the effect due to the source and observer peculiar velocities, see \cite{Cusin:2024git} for a detailed discussion. Here we assume galaxy peculiar velocities to be negligible at high redshift, and we also neglect the observer velocity, setting $\bm{\varv}_{\text{o}} \approx 0$. This corresponds to filtering out the contribution of a kinematic dipole from the source distribution function.

The source function $\mathcal{S}_i$ defines the GW energy spectrum emitted by a single binary, and it is a function of the source frame frequency $f_{\text{s},k}$ and the binary parameters $\zeta_{\text{b}}$. For BBH, it is given by
\begin{equation}\label{eq:SBBH}
    \mathcal{S}_{\text{BBH}}(f_{\text{s}},\zeta_{\text{b}})
    %\equiv \int_{S^2} \dd^2 \sigma_{\text{s}} r_{\text{s}}^2 \tilde(h)^2_{\text{BBH}}
    =\frac{5(G\,{\mathcal{M}})^{5/3}}{6\pi^{1/3}}\times\,
    \begin{cases}
        f_{\text{s}}^{-7/3}\,\Big[1+\sum_{i=2}^{3} \alpha_i\,(\pi\,G\,M\,f_{\text{s}})^{i/3}\Big]^2, &f_{\text{s}}< f_1\\
        c_1\,f_{\text{s}}^{-4/3}\, \Big[1+\sum_{i=1}^{2} \epsilon_i\,(\pi\,G\,M\,f_{\text{s}})^{i/3}\Big]^2, &f_1 \leq f_s <f_2\\
        c_2\,\Big[1+\Big(\frac{f_{\text{s}}-f_2}{f_3}\Big)\Big]^2, &f_2\le f_{\text{s}}<f_4
    \end{cases}
\end{equation}
where definitions of $c_{1,2}, \alpha_{i},\epsilon_{i}$ and $f_{1,2,3,4}$ can be found in \cite{Jenkins:2018uac, PhysRevLett.106.241101}, and depend on the masses and spins of the binaries. Here, $G$ is the Newton's gravitational constant, $M$ stands for the total mass of the binary, and $\mathcal{M}$ stands for its chirp mass. 
For BNS and BHNS systems, only $f_{\text{s}}<f_1$ terms will be kept, and contributions from higher frequencies can be ignored. 
The optimized inclination angle is assumed in this calculation.

The merger rate depicts the rate of the type $i$ compact binary mergers in galaxy $k$ and depends on the galaxy redshift, its metallicity, and the binary merger parameters 
\begin{equation}\label{eq:mergrate}
R_i(z_k,\mathcal{Z}_k,\zeta_{\text{b}})=\frac{\mathcal{R}_i^{(\text{local})}}{\mathcal{I}_i} \,p_i(\zeta_{\text{b}}) \,f_{\mathcal{Z}}\,\psi_{\text{d},i}(z_k)\,.
\end{equation}

Specifically, we assume the merger rate in a given galaxy will follow the rate $\psi$ of star-formation in the galaxy, delayed by the lifetime of the binary between its formation and merger. To capture the variability in the time delay, we average over its probability distribution. Hence, the delayed star-formation rate is given by
\begin{equation}\label{eq:psid}
    \psi_{\text{d},i} (z_k) = \frac{1}{\ln\big[t(z)/t_{\text{min},i}\big]} \int_{t_{\text{min},i}}^{t(z)} \dd t_{\text{d}}\;\frac{1}{t_{\text{d}}}\;\psi\big[z_{{\rm f}}(z_k,t_{\text{d}})\big]\,,
\end{equation}
where the delay time $t_{\text{d}}$ has a probability distribution $p(t_{\text{d}})\propto 1/t_{\text{d}}$ between the minimum delay time $t_{\text{min},i}=[20,50,50]\ \rm{Myr}$ for $i=\rm{[BNS,BBH,BHNS]}$ and the maximum delay time $t_{\text{max}}$ which is the age of the Universe $t(z)$. Here $z_{{\rm f}}$ is the formation redshift defined by the merger redshift $z_k$ and the time delay $t_{\text{d}}$ 
\be\label{eq:zform}
    1+z_{\text{f}}(z_k,t_{\text{d}})=(1+z_k)\Bigg[\cosh\bigg(\frac{3\Omega_{\Lambda}^{1/2}\,t_{\text{d}}}{2t_H}\bigg)-\frac{E(z)}{\Omega_{\Lambda}^{1/2}}\,\sinh\bigg(\frac{3\Omega_{\Lambda}^{1/2}\,t_{\text{d}}}{2t_H}\bigg)\Bigg]^{-2/3}\;,
\ee
where $E(z)\equiv H(z)/H_0=\sqrt{\Omega_{\mathrm{m}}(1+z)^3+\Omega_{\Lambda}}$ assuming standard flat $\Lambda$CDM cosmology. 
The dependence of the merger rate on metallicity is captured by the correction factor $f_{\mathcal{Z}}$. Following \cite{Jenkins:2018uac}, we assume that black holes with mass greater than $30 M_\odot$ can only be formed in galaxies with relatively small metallicity: $Z\le \frac{1}{2}Z_\odot$, or equivalently $\mathcal{Z}\le \log_{10}\frac{1}{2}\approx -0.301\equiv \mathcal{Z}_{\text{lim}}$. Consequently, 
\begin{equation}\label{eq:metallicityfunction}
    f_{\mathcal{Z}}({\mathcal{Z}},m_1,m_2)=
    \begin{cases}
        1, &m_1, m_2<30 M_\odot\,,\\
        \Theta\big(\mathcal{Z}_{\text{lim}}-\mathcal{Z}\big), &\mathrm{otherwise}\,.
    \end{cases}
\end{equation}

The dependence on the binary merger parameters $\zeta_{\text{b}}$ is captured in the normalized probability distribution $p_i(\zeta_{\text{b}})$. For simplicity, we set the neutron star spins to zero, and the masses uniformly distribute between $1.1 M_\odot$ and $2.0 M_\odot$; the black hole spins are set to be uniformly distributed between $-1$ and $1$, with a mass distribution from the fiducial Power-law + Peak (PP) model in \cite{KAGRA:2021duu} (Fig. 10). The minimum and maximum masses of black holes are $M_{\text{BH}}^{\text{min}}=5.08 M_\odot,\,M_{\text{BH}}^{\text{max}}=86.85 M_\odot$. 
For BNS, both neutron stars follow the mass distribution and spin set above. For BHNS, the neutron star and the black hole follow the mass and spin distribution set above for them respectively. For BBH, the primary black hole follows the above mass and spin distribution, while the secondary black hole in the binary system has the same spin distribution but its mass is always smaller than that of the primary black hole.

Finally, we normalize the merger rate so that its local value (i.e. at $z=0$) agrees with the fiducial model merger at redshift $z=0$ with its lowest 5\% and highest 95\% confidence interval of GWTC-3 in \cite{KAGRA:2021duu}
\begin{equation}\label{eq:localrate}
    \mathcal{R}_{\text{BBH}}^{(\text{local})}=23.9^{+14.9}_{-8.6}\ \rm{Gpc}^{-3}\ \rm{yr}^{-1},\; \mathcal{R}_{\text{BNS}}^{(\text{local})}=105.5^{+190.2}_{-83.9}\ \rm{Gpc}^{-3}\ \rm{yr}^{-1},\; \mathcal{R}_{\text{NSBH}}^{(\text{local})}=32.0^{+62.0}_{-24.0}
    \ \rm{Gpc}^{-3}\ \rm{yr}^{-1}\,.
\end{equation}
The BBH and BNS merger rates are in Sect. IV. A in \cite{KAGRA:2021duu}, and the NSBH merger rate is chosen from the BGP model in Table II in \cite{KAGRA:2021duu}. 
As we do not know the distribution of the merger rate, we have to assume it is Gaussian, and the 68\% confidence interval $\sigma$ can be approximated as 1.6 times smaller than the 90\% confidence interval in Eq. (\ref{eq:localrate}).

This requires the normalization factor $\mathcal{I}_i$ in Eq. (\ref{eq:mergrate}), which is given by~\cite{Jenkins:2018uac}
\begin{equation}\label{eq:normfactor}
    \mathcal{I}_i
    =\Bigg\{\frac{1}{\ln\big[t(z)/t_{\text{min},i}\big]}\, \int_{t_{\text{min},i}}^{t(z)}\frac{\dd t_{\text{d}}}{t_{\text{d}}}\;\psi^{(V)}(z_{\text{f}})\,\frac{n(z)}{n(z_{\text{f}})}\,\int \dd\mathcal{Z} \;p(\mathcal{Z}|z)\,\int\dd\zeta_{\text{b}}\; p_i(\zeta_{\text{b}})\,f_{\mathcal{Z}}\Bigg\}\Bigg\rvert_{z=0}\,,
\end{equation}
where $\psi^{(V)}(z)$ is the sum of the galactic star-formation rates per unit comoving volume at redshift $z$, and $n(z)$ is the number of galaxies per comoving volume. $p(\mathcal{Z}|z)$ is the probability distribution of the log-normalized metallicity $\mathcal{Z}$ over redshift $z$. We will discuss how to estimate these three quantities using catalogue information and how to compute $\mathcal{I}_i$ in detail in Sect. \ref{sec:Euclid}.

Finally, to study anisotropy in this model, the SGWB energy density is expanded into spherical harmonics, with the coefficients given by
\begin{equation}
    a_{\ell m}(f)\equiv \int_{\text{S}^2}\dd^2\sigma \; \Omega_{\text{GW}}(f,{\bf{e}})\, Y_{\ell m}^{*}({\bf{e}})
    =\sum_k \varw_k (f) \,Y_{\ell m}^{*}({\bf{e}}_k)\,.
\end{equation}

\section{The {\textit{Euclid}} Flagship simulation: galaxies}\label{sec:Euclid}

Computing the SGWB with the method described above requires a simulated galaxy catalogue; we select the \textit{Euclid} Flagship Simulation Galaxy Catalogue (v2.1.10) \cite{Euclid:2024few} hosted on CosmoHub \cite{Tallada:2020qmg,Carretero:2017zkw} because it is tailored to the properties of the \textit{Euclid} Wide Survey (EWS), thereby positioning our pipeline to later propagate and quantify survey systematics directly on real \textit{Euclid} data. This Catalogue is based on the record-setting Flagship2 $N$-body simulation that includes $\sim$4 trillion dark matter particles. It contains the properties of $\sim$4.8 billion galaxies up to $H_{\text{E}}<26.6$ considering no selection of emission lines or corrections due to Milky Way extinction, %(with no emission lines, no Milky Way extinction)
distributed over one octant of the sky ($\sim 5157\ \mathrm{deg}^2$) centered at approximately the North Galactic Pole ($145^{\circ}<\rm{RA}<235^{\circ}, 0^{\circ}<\rm{Dec}<90^{\circ}$), and in the redshift range between 0 and 3. Given that the SGWB includes signal from all galaxies, even those that are in principle not detectable, we include all simulated objects in the computation. The effects of observational selection functions in the estimate of the {\it Euclid} galaxy density and consequently in the measured cross-correlation, will be explored in a future paper.  
%The \textit{Euclid} Wide survey will nominally observe objects up to a brighter magnitude cut of $H_{\text{E}}$. We note that no observational effects specific to {\it Euclid} are considered in our analysis.

For each galaxy, the Catalogue provides the following parameters needed in our calculation using the method in Sect. \ref{sec:Eqs}: redshift $z$, star-formation rate $\psi$, stellar mass $M_*$, and metallicity $12+\log_{10}(\text{O/H})$. %, peculiar velocity $\bm{\varv}_k=(\varv_x,\varv_y,\varv_z)$, direction $\hat{\bf{e}}_k=(x,y,z)/r$.%-->we have removed peculiar velocity term in the source-frame frequency eqaution.
Following the galaxy metallicity distribution model in \cite{Curti2020} that the {\it Euclid} Flagship Simulation Galaxy Catalogue \cite{Euclid:2024few} is using, the abundance of oxygen (O) relative to hydrogen (H) is defined as $12+\log_{10}(\text{O/H})$, and the Solar value is 8.69 \cite{Allende_Prieto_2001}. The logarithmic normalized metallicity $\mathcal{Z}$ of galaxy is defined with respect to the Solar metallicity in Eq. (\ref{eq:logZ}), which is then related to the galaxy's $12+\log_{10}(\text{O/H})$ as
\begin{equation}
    \mathcal{Z}\equiv \log_{10}\frac{Z}{Z_\odot}=\log_{10}\frac{[\text{O/H}]}{[\text{O/H}]_\odot}=\log_{10}\frac{10^{[12+\log_{10}(\text{O/H})]-12}}{10^{8.69-12}}=\big[12+\log_{10}(\text{O/H})\big]-8.69\,.
\end{equation}
%The cosmological parameters used in the $N$-body dark matter simulation are: $\Omega_{\rm{m}} = 0.319, \Omega_{\rm{b}} = 0.049, \Omega_\Lambda + \Omega_\gamma = 0.681, A_s = 2.1\times10^{-9}, n_s = 0.96, h = 0.67$, and the particle mass is $m_p\sim 1.0\times10^{9} M_\odot/h$.
%The catalogue includes:spectroscopic and photometric information, lensing properties and shape parameters.

%\subsection{galaxy star-formation history model}
In order to compute the merger rate in Eq. (\ref{eq:mergrate}), our model requires the galaxy's delayed star-formation rate $\psi_{\rm{d}}$. This, in turn, requires a model of the galaxy SFH, which is not provided by the {\textit{Euclid}} Flagship Simulation Galaxy Catalogue. We therefore impose an SFH model that leverages the information provided by the Catalogue: stellar mass $M_*$ and star-formation rate $\psi(z_{\text{obs}})$ of the galaxy at its current redshift $z_{\text{obs}}$. 

Following the literature on the topic \cite{Renzini:2006je,Thomas_2005,Mason_2015,Vangioni:2014axa,Madau2014}, we implement a mixed model. Specifically, galaxies of stellar mass $M_*>10^{10} M_\odot$ follow a log-normal exponential SFH model \cite{Diemer:2017ttl}, given by
\begin{equation}
   \psi(t)=\frac{A}{\sqrt{2\pi \tau^2}\;t}\,\exp \Bigg\{-\frac{\Big[\ln\big(\frac{t}{1\,\mathrm{Gyr}}\big)-T_0\Big]^2}{2\tau^2}\Bigg\}\,,
\end{equation}
where $A, T_0, \tau$ are parameters, $t$ is the time since the Big Bang. The cumulative SFH $\psi_{\rm{c}}$, which is the integral of the SFR, gives the current stellar mass $M_*$ of the galaxy at the time of observation
\begin{equation}
    \psi_{\rm{c}}(t)\equiv \int_{0}^{t} \dd t^{\prime}\; \psi(t^{\prime})=\frac{A}{2}\,\Bigg\{1-\text{erf} \bigg[-\frac{\ln\big(\frac{t}{1\,\mathrm{Gyr}}\big)-T_0}{\tau\sqrt{2}}\bigg]\Bigg\}\,,
\end{equation}
\begin{equation}
    \psi_{\rm{c}}(t_{\text{obs}})=10^9\,M_*\,.
\end{equation}
In order to constrain the three parameters $A, T_0, \tau$ with the two known quantities $\psi$, $M_*$, the power-law approximation of the relation between the size and centre of the $\psi-t$ curve peak is needed \cite{Diemer:2017ttl}
\begin{equation}\label{eq:t-sigma}
    \sigma_{\text{peak}}=0.83\, t_{\text{peak}}^{3/2}\,,
\end{equation}
These two peak parameters also depend on the three parameters mentioned above \cite{Diemer:2017ttl}
\begin{equation}
    \frac{t_{\text{peak}}}{1\,\mathrm{Gyr}}={\mathrm{e}}^{T_0-\tau^2},\; \sigma_{\text{peak}}=2\,\frac{t_{\text{peak}}}{1\,\mathrm{Gyr}}\,\sinh\Big[\sqrt{2\ln(2)}\,\tau\Big]\,.
\end{equation}
The power-law relation in Eq. (\ref{eq:t-sigma}) is valid for a small portion of galaxies, so not all galaxies have numerical solutions using the log-normal exponential model. For those galaxies without valid solutions in this model, especially less massive galaxies, their $\sigma_{\text{peak}}$ is large enough that it is reasonable to treat their SFH as flat
\begin{equation}
    \psi(t)=\psi(t_{\text{obs}})\;
\end{equation}
between a start time $t_0$ and the time the galaxy is observed $t_{\text{obs}}$. So the integral of the SFH gives the stellar mass $M_*$
\begin{equation}
    \int_{t_0}^{t_{\text{obs}}} \dd t^{\prime}\;\psi(t^{\prime})= (t_{\text{obs}}-t_0)\, \psi(t_{\text{obs}})=M_*\,,
\end{equation}
hence the start time $t_0=t_{\text{obs}}-M_*/\psi(t_{\text{obs}})$.

\begin{figure}[!ht]
    \centering
    \includegraphics[width=0.65\linewidth]{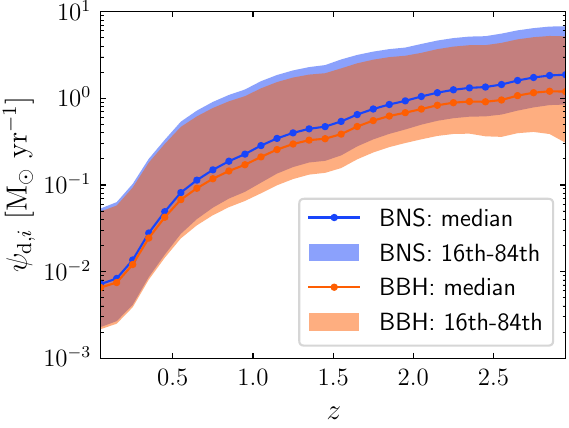}
    \caption{The median, 16th and 84th percentile of the delayed star-formation rate computed for minimum time $20$ and $50\ \rm{Myr}$ (corresponding to BNS and BBH or BHNS respectively) using the exponential-flat mixed SFH model for all {\textit{Euclid}} Flagship Simulation Galaxy Catalogue galaxies with $H_{\text{E}}$ less than 26.6 in redshift bins of 0.1 width.}
    \label{fig:galpsid}
\end{figure}

The galaxy delayed star-formation rate is defined in Eq. (\ref{eq:psid}). For some galaxies, the time since the first stars formed is shorter than $50\ \rm{Myr}$, so they are not able to form BBH or BHNS ($\psi_{\text{d,BBH}}=\psi_{\text{d,BHNS}}=0$). On the other hand, all {\textit{Euclid}} Flagship Simulation Galaxy Catalogue galaxies have SFH longer than $20\ \rm{Myr}$, so they are all able to form BNS. The median values with the 16th and 84th percentile of the distribution of $\psi_{\text{d},i}$ for Catalogue galaxies in redshift bins of width 0.1 between 0 and 3 are shown in Fig. \ref{fig:galpsid}.

%\subsection{Evaluation of rate function normalization factor}\label{subsec:Ii}
Further, the calculation of the merger rate in Eq. (\ref{eq:mergrate}) also requires the normalization factor $\mathcal{I}_i$, which in turn relies on estimates of the redshift-dependent galaxy number density $n(z)$, the star-formation rate density $\psi^{(V)}(z)$, and the log-normalized metallicity probability distribution $p(\mathcal{Z}|z)$, as specified in Eq. (\ref{eq:normfactor}). Recall from Eq. (\ref{eq:metallicityfunction}) that $f_{\mathcal{Z}}$ is a selection function such that only galaxies with $\mathcal{Z}\le \mathcal{Z}_{\text{lim}}$ can produce black holes with mass $\ge 30 M_\odot$. The last two integrals of $\mathcal{I}_i$ in Eq. (\ref{eq:normfactor}) for $i$=BNS are unity; for $i$=BBH or BHNS, they can be written as
\begin{equation}
    \begin{split}
    \int_{-\infty}^{\mathcal{Z}_{\text{max}}}{\mathrm{d}}\mathcal{Z} \; p(\mathcal{Z}|z) \int {\mathrm{d}}\zeta_{\text{b}}\;
    p_i(\zeta_{\text{b}})\,f_{\mathcal{Z}}
    &=\Bigg[ p\Big(\mathcal{Z}<\mathcal{Z}_{\text{lim}}|z\Big) \int_{M_{\text{BH}}^{\text{min}}}^{M_{\text{BH}}^{\text{max}}}\dd m_1
    +p\Big(\mathcal{Z}\ge\mathcal{Z}_{\text{lim}}|z\Big) \int_{M_{\text{BH}}^{\text{min}}}^{30 M_\odot}\dd m_1\Bigg]\\
    &\times 
    \begin{cases}
    \begin{aligned}
        \int_{M_{\text{BH}}^{\text{min}}}^{m_1}\dd m_2 \int_{-1}^{1}\dd \chi_{1}\int_{-1}^{1}\dd \chi_{2} \; p_i(\zeta_{\text{b}}) \,,\quad&\text{BBH}\\
        \noalign{\vskip9pt}
        \int_{1.1 M_\odot}^{2.0 M_\odot}\dd m_2 \int_{-1}^{1}\dd \chi_1\; \delta(\chi_2)\, p_i(\zeta_{\text{b}}) \,,\quad&\text{BHNS}
    \end{aligned}
    \end{cases}
    \end{split}\;.
\end{equation}
Using information provided in the {\it Euclid} Flagship Simulation Galaxy  Catalogue, we compute the three quantities $n(z)$, $\psi^{(V)}(z)$, and $p(\mathcal{Z}<\mathcal{Z}_{\text{lim}}|z)$ in redshift bins of width 0.01\footnote{Note that no observational effects specific to {\it Euclid} are included in this computation.}. %, and show the results in Fig. \ref{fig:Ii}. \textcolor{red}{The SFR density and metallicity distributions show discrete steps in redshift by step 0.5, which is a result of how the \textit{Euclid} simulation is performed (model of H$\alpha$ flux).} 
We then use these distributions and a local estimation of $n(z=0)\approx 0.223\ \mathrm{Mpc}^{-3}$ to compute $\mathcal{I}_i$ in Eq. (\ref{eq:normfactor}) for all types of binaries (BNS, BBH, BHNS).
%\begin{figure}[!ht]
%    \centering
%    \includegraphics[width=.3\textwidth]{plots/NbarV_z.pdf}
%    \hfill
%    \includegraphics[width=.3\textwidth]{plots/psibarV_z.pdf}
%    \hfill
%    \includegraphics[width=.3\textwidth]{plots/PZ_z.pdf}
%    \caption{(Left:) number density (center:) star-formation rate density in unit comoving volume (right:) probability distribution of log-normalized metallicity smaller than $\mathcal{Z}_{\text{lim}}$ at redshift z for galaxies in the {\textit{Euclid}} catalogue.}
%    \label{fig:Ii}
%\end{figure}

%\subsection{{\textit{Euclid}} catalogue limits}
Next, we note that there are three observational limitations of the {\textit{Euclid}} Flagship Simulation Galaxy Catalogue. First, the Catalogue covers only $f_{\text{sky}}=1/8$ of the sky. This limit can be justified by evaluating all the galaxy densities properties (i.e., number density, star-formation rate density, etc.) within $f_{\text{sky}}$ of the comoving volume and scaling up their density per comoving value by $1/f_{\text{sky}}$. The angular power spectra $C_\ell$ should also scale up by $1/f_{\text{sky}}$. This scaling is necessary for any galaxy catalogue that has partial-sky coverage. 

\begin{figure}[!ht]
    \centering
    \includegraphics[width=0.46\linewidth]{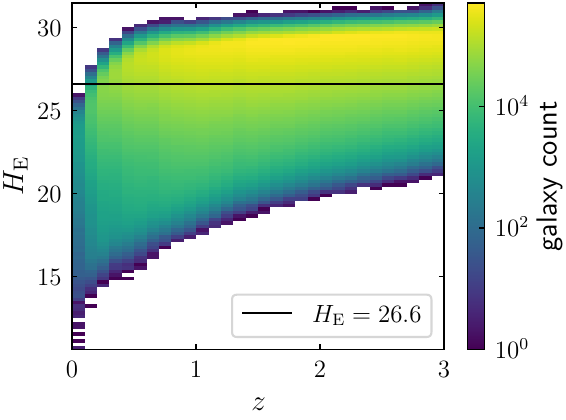}
    \hfill
    \includegraphics[width=0.48\linewidth]{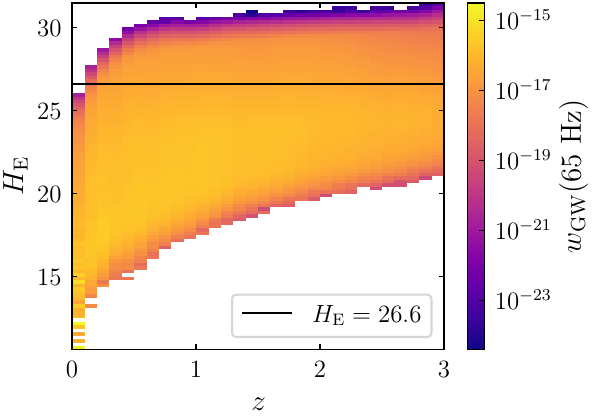}
    \caption[{\it Euclid} deep region]{2D histogram of $H_{\text{E}}$ and redshift with the colour bar being galaxy number count (\emph{left}) or gravitational wave energy density (\emph{right}) of all galaxies in {\it Euclid} deep region in $5^{\circ}<\rm{Dec}<10^{\circ}, 150^{\circ}<\rm{RA}<155^{\circ}$ with redshift up to 3. The black lines stand for $H_{\text{E}}=26.6$.}
    \label{fig:deepregion}
\end{figure}

Second, the Catalogue contains only galaxies with $H_{\text{E}}$ less than 26.6. However, it also includes a $5\times5\ \rm{deg}^2$ deep region ($5^{\circ}<\rm{Dec}<10^{\circ}, 150^{\circ}<\rm{RA}<155^{\circ}$) that contains all galaxies with redshift between 0 and 3, and without any $H_{\text{E}}$ magnitude constraint. This deep region indicates that a large portion ($80\%$) of galaxies has $H_{\text{E}}$ greater than or equal to 26.6. Nevertheless, due to their relatively large distance from Earth, these galaxies contribute little to the total GW energy density, less than $5\%$. Therefore, we ignore it for the rest of our calculation as it does not qualitatively impact our results. The distribution of galaxy number counts and GW energy density contributions as 2D histograms of $H_{\text{E}}$ and redshift are shown in Fig. \ref{fig:deepregion}. 

Third, the Catalogue contains only galaxies with a redshift smaller than 3. Under the assumption that most of the star-formation takes place at redshifts below 3 \cite{Madau2014}, combined with the stronger distance suppression of GW signals from higher redshifts, (cf. Eq. \ref{eq:OGW}), the SGWB energy density will be dominated by contributions from redshifts 2 and below. The redshift cutoff will, therefore, not have a qualitative impact on our results, and we demonstrate this in Sect. \ref{subsec:totalOGW}.

\section{Estimates of the Gravitational-Wave Background and its Angular Power Spectra}\label{sec:PredCBC}

\subsection{Total Gravitational-Wave Energy Density}\label{subsec:totalOGW}

Using data from the {\textit{Euclid}} Flagship Simulation Galaxy Catalogue and following the procedure outlined in Sections \ref{sec:Eqs} and \ref{sec:Euclid}, we compute the average expected GW energy density contribution $\varw_k$ (cf. Eq. \ref{eq:OGW}) for every {\textit{Euclid}} Flagship galaxy and for each of the three binary types: BBH, BNS, and BHNS. We emphasize that we include all galaxies from the Catalogue in this calculation, without including any observational effects specific to {\it Euclid}. Integrating over all of the galaxies in the Catalogue (and accounting for the factor of 8 due to the partial sky coverage) yields a prediction for the total GW energy density due to compact binary mergers. 

Table \ref{tab:OGWtot} shows these predictions in comparison to the predictions of the semi-analytic model based on the latest version of the GW catalogue, GWTC-3 \cite{KAGRA:2021duu}. Our predictions for the energy density are about 10 times smaller than those of the semi-analytic model. We note that the semi-analytic model assumes the merger rate to follow the global star-formation rate in the Universe, convolved with the formation-to-merger time delay distribution. This is different from the galaxy-catalogue-based approach, which is limited to the galaxies available in the catalogue, and handles SFH separately for each individual galaxy. 
%{\color{red} CS: but does this explain the factor of ten? I thought that the GW power predicted for the deepest part of the FS catalogue was not a x10 different. }

The paper by Jenkins and others~\cite{Jenkins:2018uac} has a calculation similar to the catalogue-based approach using the {\it Millennium} simulation. They do not give predictions of the total GW energy density, as they used a restricted sample of galaxies with $z<0.78$. However, they give an analytical prediction that is 10 times bigger than the semi-analytic model. We note that this analytical calculation was based on earlier estimates of binary merger rates that were higher than those used both in this work and in the GWTC-3 estimate, being $\mathcal{R}_{\text{BBH}}^{(\text{local})}=103^{+110}_{-63}\,\rm{Gpc}^{-3}\,\rm{yr}^{-1},\;\mathcal{R}_{\text{BNS}}^{(\text{local})}=1540^{+3200}_{-1220}\,\rm{Gpc}^{-3}\,\rm{yr}^{-1},\; \mathcal{R}_{\text{NSBH}}^{(\text{local})}\le 3600\,\rm{Gpc}^{-3}\,\rm{yr}^{-1}$, where the BBH and BNS rates are from implications of SGWB from the GW170817 event \cite{GW170817implySGWB}, and the BHNS rate is an upper limit from LIGO O1 \cite{O1rateUL}. As we continue to observe more GW transients through observing runs, the estimate of merger rates has improved. 

Finally, we note that our predictions are consistent with the latest 95\% confidence upper limit on GW background energy density, $\Omega_{\text{GW}}(\text{25\ Hz})\le 2.0\times 10^{-9}$ for a power-law GW background with a spectral index $\alpha=2/3$ using data from the first part of LIGO, Virgo, and KAGRA’s fourth observing run~\cite{O4a-isoback}. In our model presented in Sect. \ref{sec:Eqs}, as we can see from Eq. (\ref{eq:OGW},\ref{eq:SBBH}), our prediction for $\Omega_{\text{GW}}$ also follows a power-law dependence on frequency $\propto f^{2/3}$ for BNS, BHNS, and BBH within the inspiral stage. 
\begin{table}[!ht]
    \centering
    \begin{tabular}{|c|c|c|c|c|c|}
    \hline
    Model & $\Omega_{\text{GW}}^{\text{tot}}(65\,{\text{Hz}})$ & $\Omega_{\text{GW}}^{\text{BNS}}(65\,{\text{Hz}})$& $\Omega_{\text{GW}}^{\text{BBH}}(65\,{\text{Hz}})$ &$\Omega_{\text{GW}}^{\text{NSBH}}(65\,{\text{Hz}})$\\
    \hline
    {\it Euclid}&$1.73^{+1.67}_{-0.82}\times10^{-10}$ &$2.15^{+3.88}_{-1.71}\times10^{-11}$ &$1.25^{+0.78}_{-0.45}\times10^{-10}$ &$2.57^{+4.99}_{-1.93}\times10^{-11}$\\
    \hline
    GWTC-3&$1.3^{+0.6}_{-0.4}\times10^{-9}$&$1.1^{+3.2}_{-0.9}\times10^{-10}$&$9.4^{+2.6}_{-3.4}\times10^{-10}$&$1.7^{+4.2}_{-1.3}\times10^{-10}$\\
    %\hline
    %     {\it Millennium}&$1.1\times 10^{-8}$&$1.2\times10^{-9}$&$1.5\times10^{-9}$&$1\times10^{-8}$\\
    \hline
    \end{tabular}
    \caption{Predicted total GW energy density for all three types of binaries (BNS, BBH, BHNS) with their highest 95\% and lowest 5\% confidence intervals from various methods. See text for details.}
    \label{tab:OGWtot}
\end{table}

\begin{figure}[!ht]
    \centering
    \includegraphics[width=0.48\linewidth]{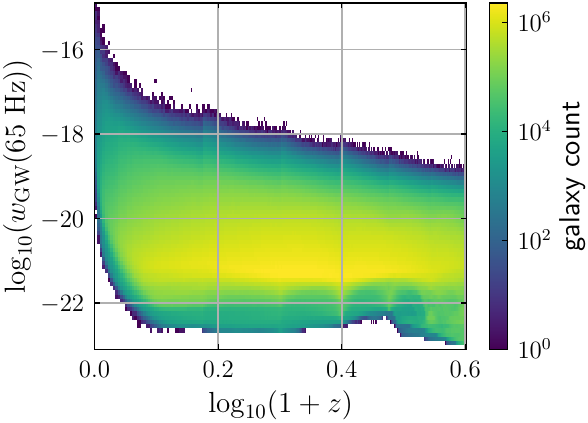}
    \hfill
    \includegraphics[width=0.475\linewidth]{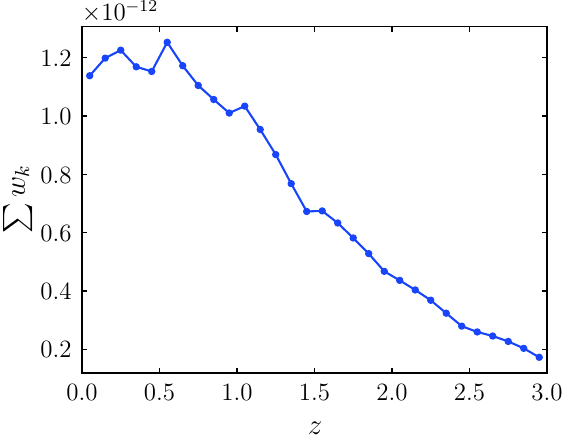} 
    \caption{\emph{Left}: 2D binned histogram of $\log_{10}{\varw^{\text{GW}}_k}$ at 65 Hz and its $
    \log_{10}(1+z)$ for all {\textit{Euclid}} galaxies. The colour bar shows number of galaxies in each bin. \emph{Right}: sum of total GW energy density in 0.1-wide redshift bins from 0 to 3.}
    \label{fig:ogw_z}
\end{figure}

To assess the contributions to the total GW energy density across galaxies, Fig. \ref{fig:ogw_z} shows a 2D histogram of $\varw^{\text{GW}}_k$ at 65 Hz as a function of the galaxy redshift, binned in 0.1-wide redshift bins. The left panel of Fig. \ref{fig:ogw_z} shows that most of the galaxies contributing to the total GW energy density have $\varw^{\text{GW}}_k \sim 10^{-22}-10^{-20}$, with galaxies at the lowest redshifts can contribute as much as $\varw^{\text{GW}}_k \sim 10^{-16}$ (although these very loud galaxies are relatively rare). Summing over all galaxies in a given redshift bin, the right panel of Fig. \ref{fig:ogw_z} shows that most of the GW energy density comes from relatively low redshifts ($z \lesssim 1$), consistent with past work \cite[e.g.,][]{wu_cbc}. This figure also shows that the sum of GW energy density at redshift $z=3$ is about 15\% of that at $z=0.1$. Coupling with the absence of significant star-formation at higher redshifts, this result confirms that the redshift cut of $z<3$ does not have a significant impact on our estimates.

\subsection{Gravitational-Wave Background Anisotropy}\label{subsec:GWanisotropy}

\begin{figure}[!ht]
    \centering
    \includegraphics[width=0.65\linewidth]{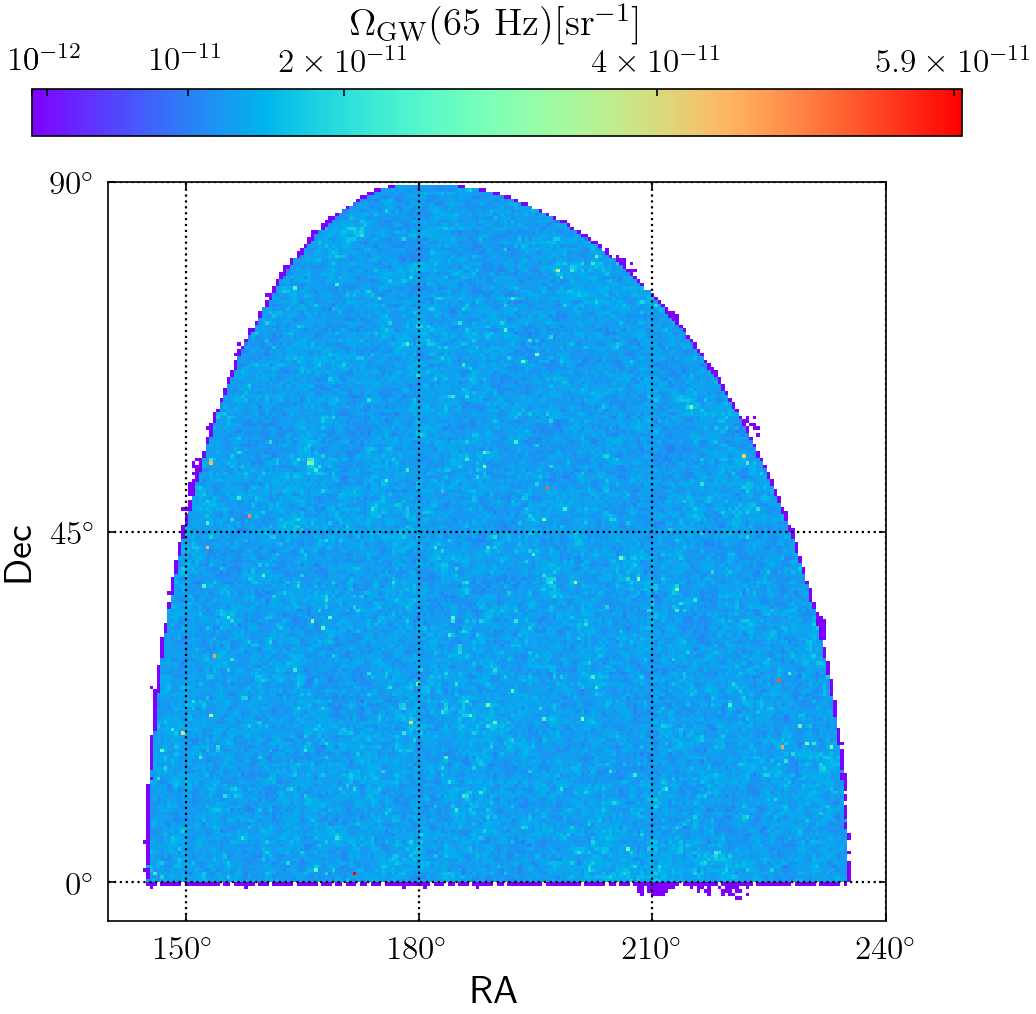}
    \caption{Total GW energy density summed over all types of binaries (BNS, BBH, BHNS) at 65 Hz for all {\textit{Euclid}} Flagship galaxies with $H_{\text{E}}$ less than 26.6 per angular size in \texttt{HEALPix} for pixels of $N_{\text{pix}}=786\,432$ with removal of the edge pixels. The colours are on a log scale. The pixels in purple stand for the edge of the sky coverage, and are removed in all following calculation.}
    \label{fig:ogw_hp}
\end{figure}

To study the anisotropy in our Catalogue-based model, we divide the sky into $N_{\mathrm{pix}} = 786\,432$ pixels of 0.052 deg$^2$ each in \texttt{HEALPix} \cite{Zonca2019,HEALPix} basis, equivalently $N_{\text{side}}=256$. We choose such a high resolution to search for the small scale structure so as to compare it to the predictions of the astrophysical models in Sect. \ref{Sec:TheoryModel}. We then sum the GW energy density contributions from all galaxies in a given angular pixel $p$, normalizing it to the angular area of the pixel
\begin{equation}
    \hat \Omega_p(f) = \frac{N_{\mathrm{pix}}}{4\pi}\,\sum_{k \in p} \sum_i \varw_{k,i}(f)\,. 
\end{equation}
$\Omega_p$ can then be used to display the GW energy density sky map using \texttt{HEALPix} \cite{Zonca2019,HEALPix}. Since pixels at the edge of the sky region covered by the Catalogue are not fully populated by galaxies, these pixels (2023 out of 999\,88 pixels) are removed from our analysis, as shown in Fig. \ref{fig:ogw_hp}. Consequently, the sky coverage fraction is modified to $1/f_{\text{sky}}=8.028$. 

%\subsection{GW auto-correlation}
To compute the angular power spectra, we first define the pixel fluctuations in GW energy density relative to the average pixel energy density: $\delta\hat\Omega_p=\hat\Omega_p-\bar{\Omega}$. Here, the average pixel value $\bar{\Omega}$ is computed over the fully populated pixels covered by the {\it Euclid} Flagship Simulation Galaxy Catalogue. This pixelized sky map is then converted into a spherical harmonics decomposition, following \cite{Zonca2019}
\begin{equation}\label{eq:gwalm}
    \hat{a}_{\ell m}=\frac{4\pi}{N_{\mathrm{pix}}}\,\sum_{p=0}^{N_{\mathrm{pix}}-1}\,Y_{\ell m}^{*}({\bf e}_p)\, \delta \hat\Omega_p\,,
\end{equation}
where ${\bf e}_p$ is a unit vector pointing at the center of pixel $p$ with sky location $(\theta,\phi)$.

\begin{figure}[!ht]
    \begin{center}
    \includegraphics[width=0.65\linewidth]{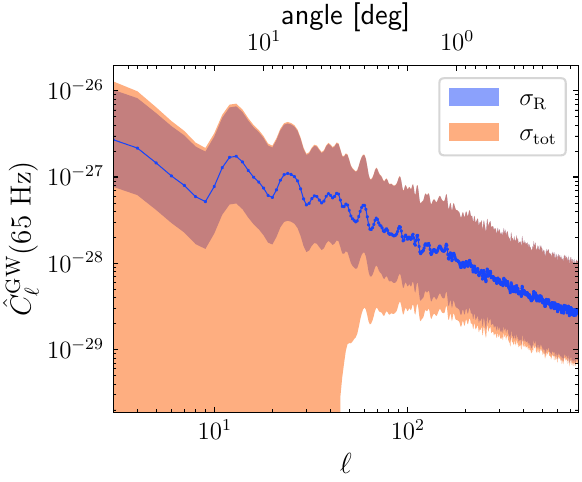}
    \caption{GW energy density fluctuation $C_\ell$ defined in Eq. (\ref{eq:gwcl}) is shown as a solid line, including contributions from all three binary types (BBH, BNS, and BHNS). Two 90\% confidence uncertainty regions are shown: the blue region shows the uncertainty due to the uncertainty in the local merger rate, while the orange region shows the total uncertainty in Eq. (\ref{eq:totsigmaClGW}).}
    \label{fig:gwcl}
    \end{center}
\end{figure}

Following Sect. \ref{sec:Eqs} and using the {\textit{Euclid}} Flagship Simulation Galaxy Catalogue, we compute the expected GW background angular power spectrum due to compact binary mergers\footnote{Again, we note that no observational effects specific to {\it Euclid} are included in this calculation.}. 
The angular power spectrum $\hat C_\ell^{\mathrm{GW}}$ is then estimated using the fluctuations as
\begin{equation}\label{eq:gwcl}
    \hat C_\ell^{\mathrm{GW}}=\frac{1}{(2\ell+1)\,f_{\text{sky}}}\,\sum_{m = -\ell}^{\ell}\,\big\lvert \hat{a}_{\ell m} \big\rvert ^2\,,
\end{equation}
with the corresponding cosmic variance being
\begin{equation}
\label{Eq:CosmicVariance}
    \sigma^2_{\text{C}}\big(\hat C_\ell^{\mathrm{GW}}\big)= \frac{2\, \big(\hat C_\ell^{\mathrm{GW}}\big)^2}{(2\ell+1)\,f_{\text{sky}}}\,.
\end{equation}
Another source of variance comes from the uncertainty in the local merger rate in Eq. (\ref{eq:localrate}). 
This contribution $\sigma^2_{\text{R}}\big(\hat{C}_\ell^{\text{GW}}\big)$ is estimated from the variations in $\hat{C}_\ell^{\rm GW}$ obtained by varying the merger rates of different types of binaries (BNS, BBH, BHNS) under the assumption of a Gaussian distribution.
%This contribution $\sigma^2_{\text{R}}\big(\hat{C}_\ell^{\text{GW}}\big)$ can be estimated by the difference between $\hat{C}_\ell^{\text{up}}$ and $\hat{C}_\ell^{\text{low}}$, computed from the GW map with the GW energy density of different types of binaries (BNS, BBH, BHNS) using the corresponding 95\% upper bound and 5\% lower bound of the local merger rate in Eq. (\ref{eq:localrate}), converting to 1 $\sigma$ by assuming a Gaussian distribution.{\color{red}{Kate: remove Eq. (\ref{eq:sigmaRClGW}}) and (\ref{eq:cosmicvar12}) ?}%from the upper (95\%) and lower (5\%) bounds of the merger rate as %the variance from the upper (95\%) and lower (5\%) bound of the local merger rate can be approximated by $1/1.645$ of half the range of the $\hat{C}_\ell^{\text{GW}}$:
%\begin{equation}\label{eq:sigmaRClGW}
%\sigma^2_{\text{R}}\big(\hat{C}_\ell^{\text{GW}}\big)= \bigg(\frac{\hat{C}_\ell^{\text{up}}-\hat{C}_\ell^{\text{low}}}{2\cdot1.6}\bigg)^2\,.
%\end{equation}
%where $\hat{C}_\ell^{\text{up}}$ (or $\hat{C}_\ell^{\text{low}}$) is computed from the GW map with the GW energy density of different types of binaries (BNS, BBH, BHNS) using the corresponding 95\% upper bound (or 5\% lower bound) of the local merger rate in Eq. (\ref{eq:localrate}). 
The total variance is then given by the two variances added quadratically
\begin{equation}\label{eq:totsigmaClGW}
    \sigma^2_{\text{tot}}\big(\hat{C}_\ell^{\text{GW}}\big)=\sigma^2_{\text{C}}\big(\hat{C}_\ell^{\text{GW}}\big)+\sigma^2_{\text{R}}\big(\hat{C}_\ell^{\text{GW}}\big)\,.
\end{equation}
The result is shown in Fig. \ref{fig:gwcl}, including the uncertainty due to cosmic variance and due to the uncertainty in the local merger rates given in Eq. (\ref{eq:localrate}). We can see from this figure that the total variance $\sigma^2_{\text{tot}}$ is dominated by the uncertainty of the local merger rate $\sigma^2_{\text{R}}$ at high $\ell$, but for $\ell<20$, it is dominated by cosmic variance $\sigma^2_{\text{C}}$, and its lower bound lies below the visible range on the logarithmic scale. This is because the partial sky coverage correction $f_{\rm sky}$ appears in the definition of cosmic variance in Eq. (\ref{Eq:CosmicVariance}), implying that cosmic variance is of similar magnitude to $(\hat{C}_\ell^{\text{GW}})^2$ at low $\ell$. The figure also reveals oscillatory features in the $\hat{C}_\ell^{\text{GW}}$ as a function of $\ell$, primarily resulting from the fact that our sky coverage (roughly one octant of the sky) cannot provide information on large angular scales (low multipoles). To mitigate these fluctuations at higher $\ell$, the curve can be smoothed by averaging over adjacent multipole bins rather than evaluating each integer $\ell$ individually.

%The predicted $C_\ell$  are  $\approx 10^{10}$ times lower than the latest observational LIGO upper limits in broadband (frequency 20--1000 Hz)  $C_\ell$ derived from the LIGO O3 observing run data \cite{KAGRA:2021mth} for the power-law index $\alpha = 2/3$. We also compare to $\hat{C}_\ell^{\text{GW}}$ in narrow frequency band 60--70 Hz in \cite{Yang:2023eqi} Fig. 4, which is dominated by LIGO detector noise, and is around $10^{-14}$ for $\ell_{\text{max}}=5$, being $10^{12}$ times larger than our result in Fig. \ref{fig:gwcl}.

Our predicted $C_\ell$ are $\sim 10^{10}$ below the latest broadband (20--1000~Hz) LIGO O3 \emph{upper limits} on $C_\ell$ \cite{KAGRA:2021mth}, which are derived assuming a power-law index \(\alpha=2/3\).  We also compare with the narrow-band (60--70\,Hz) LIGO upper limit of $\hat{C}_\ell^{\mathrm{GW}}$ reported in Fig.~4 of \cite{Yang:2023eqi} ($\hat{C}_\ell^{\mathrm{GW}}\sim 10^{-14}$ at $\ell_{\max}=5$).  Similarly to $C_\ell$, the current upper limits are well  above the predicted signal, due to the limited sensitivity of the LIGO O3 campaign.

\begin{figure}[!ht]
    \centering
    \includegraphics[width=0.65\linewidth]{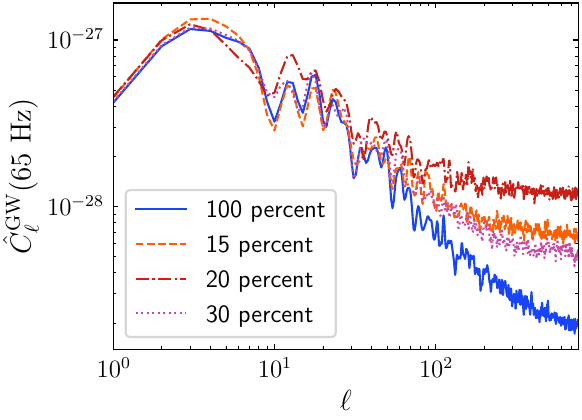}
    \caption{GW angular power spectra at 65 Hz using all (100\%) or 15\%, 20\%, 30\% of galaxies with $0<z<0.1$ in the {\textit{Euclid}} Flagship Simulation Galaxy Catalogue. The angular spectra computed with fewer galaxies plateau at high values of $\ell$, and the plateau starts at lower values of $\ell$ if there are fewer galaxies in the sample.}
    \label{fig:sampling}
\end{figure}

%subsubsection{shot noise effect}
We note that estimates of the GW angular power spectra may be shot-noise-limited if the number of available galaxies in the catalogue is small. To illustrate this effect, we focus on the redshift bin $0<z<0.1$, in which the {\it Euclid} Flagship Simulation Galaxy Catalogue contains about 8.5 million galaxies (out of 4.8 billion in the entire Catalogue). The galaxies are split into $786\,432$ \texttt{HEALPix} sky map pixels of size 0.052 deg$^2$. We then apply the procedure developed in Sect. \ref{sec:Eqs} to compute the SGWB angular power spectrum $\hat C_\ell^{\mathrm{GW}}$ defined in Eq. (\ref{eq:gwcl}), which is shown as the blue solid curve in Fig. \ref{fig:sampling}. We then randomly sample 15\%, 20\%, 30\% of galaxies in each \texttt{HEALPix} pixel, and repeat this calculation. As galaxy count in pixels must be integers, the above choices of percentages ensure that every valid pixel is included during sampling (i.e., sample number in pixel $\ge 1$), so we are using the same pixels without sampling for all the choices of samples. For every choice of samples, we replace the average of the GW energy density of all galaxies in every pixel by the average of the sampled galaxies. The resulting $\hat{C}_\ell^{\text{GW}}$ is shown in Fig. \ref{fig:sampling}. As we can see in Fig. \ref{fig:sampling}, when computed with fewer galaxies, $\hat{C}_\ell^{\text{GW}}$ plateaus at high values of $\ell$ as compared to the no-sampling case when all galaxies are used, and the plateau starts at lower $\ell$ as the sample size decreases. Hence, care must be taken when computing the angular power spectra to avoid possible bias due to shot noise limitations. Please note that the $\hat{C}_\ell^{\text{GW}}$ with 15\% sampling is at lower level at $\ell\ge 10^2$ compared to 20\% sampling, this is possibly due to the loss of energy with an incomplete catalogue. Nevertheless, this does not affect the trend that larger size of samples plateaus at higher values of $\ell$, as the 20\% sampling curve plateaus at higher values of $\ell$ compared to the 15\% sampling curve. 
We note that Fig. 4 of \cite{Jenkins:2018uac} exhibits similar behavior -- since this study was done using the {\it Millennium} simulation with about 5.7 million galaxies, we speculate that high-$\ell$ limit of their catalogue-based approach may have been impacted by the shot noise limitation. 
%shot noise effect rises when the galaxy catalogue contains limited number of galaxies. The {\textit{Euclid}} Flagship simulation catalogue has fewest galaxies with redshift $<0.1$, which is only about 8.5 million, while the complete catalogue has 4.8 billion galaxies of $0<z<3$. We have a further study of galaxies within this range and find that with fewer galaxies over the sky, higher $\ell$ information is lost. 
%We use the computed GW energy density of all {\textit{Euclid}} galaxies with $0<z<0.1$, then randomly pick 2\%, 5\% and 10\% samples in each \texttt{HEALPix} sky map pixel, finally compute the angular power $C_\ell$s of the sampled GW energy density map. The result is shown in Fig. \ref{fig:sampling}. As we can read from the figure, with sampling, $C_\ell$ vs. $\ell$ becomes flat at $\ell$ between 10 to 100. Meanwhile, the fewer samples are used, $C_\ell$ vs. $\ell$ become flat at lower $\ell$. However, if we use all the galaxies without sampling, the $C_\ell$ vs. $\ell$ keeps decreasing at $\ell$ between 100 to 1000. This clearly demonstrates the shot noise effect.

%When we compare to the GW auto energy density $C_\ell$s at 65.75 Hz in Figure 4 of \cite{Jenkins:2018uac}, we can see that their catalogue approach $C_\ell$s are smaller than their analytical approach $C_\ell$s from $\ell \ge 10$. On this $\ell(\ell+1)C_\ell/(2\pi)$ vs. $\ell$ figure, their catalogue approach $C_\ell$s drop by one magnitude compare to their analytical approach $C_\ell$s at high $\ell$. Once we convert this figure into $C_\ell$ vs. $\ell$, the catalogue approach curve become flat for $\ell\ge 100$. This trend clearly presents the shot noise effect when the galaxy catalogue does not contain enough number of objects as we stated above. 

\subsection{Cross-Correlating GW Background and Galaxy Distribution}\label{subsec:cross-corr}

We next compute the angular power spectrum for cross-correlation between the GW background and the galaxy count distribution in the {\textit{Euclid}} Flagship Simulation Galaxy Catalogue. As discussed above, this angular spectrum can be used as another astrophysical and cosmological probe \cite{Cusin:2019jpv}.

The spherical harmonic decomposition of the GW background is discussed in Sect. \ref{subsec:GWanisotropy}. Regarding the galaxy counts, we define $\hat n_p$ as the number of Catalogue galaxies in the sky pixel $p$, and then compute the fluctuations in the galaxy number as $\hat \Delta_p = (\hat n_p-\bar{n})/\bar{n}$, where $\bar{n}$ is the average number of galaxies per pixel. The spherical harmonic decomposition is then defined as
\begin{equation}
    \hat{b}_{\ell m}=\frac{4\pi}{N_{\mathrm{pix}}}\,\sum_{p=0}^{N_{\mathrm{pix}}-1}\,Y_{\ell m}^{*}({\bf e}_p)\, \hat \Delta_p\,.
\end{equation}

\begin{figure}[!ht]
    \begin{center}
    \includegraphics[width=0.65\linewidth]{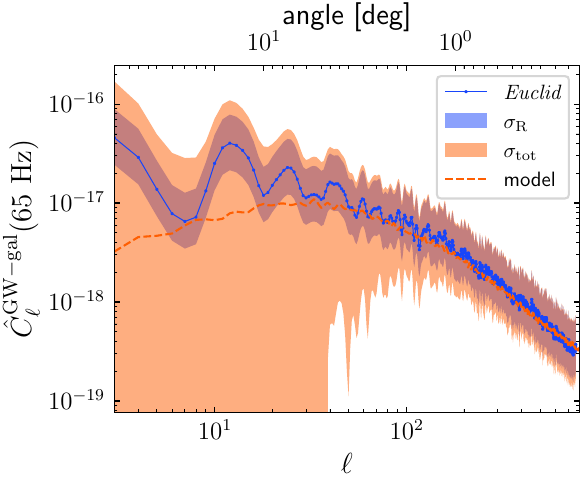}
    \caption{Angular spectrum for cross-correlation between GW background and galaxy over-density is shown as a solid blue line. Two 90\% confidence uncertainty regions are also shown including the uncertainty due to the local binary merger rate (blue) and due to the total variance (orange). For comparison, we show the red dashed line which is the astrophysical model $C_{\ell}^{\text{cross}}(\theta)$ in Eq. (\ref{eq:modelCl}) with the parameters $\theta$ being $A_0=1.42\times10^{-37}\text{erg\,cm}^{-3}\,\text{s}^{-1/3},\,z_c=1.08,\,\sigma_z=0.27$.}
    \label{fig:crosscl}
    \end{center}
\end{figure}

The cross-correlation angular power spectrum is then computed as
\begin{equation}\label{eq:crosscl}
    \hat{C}_\ell^{\text{GW-gal}}=\frac{1}{(2\ell+1)\,f_{\text{sky}}}\, \sum_{m=-\ell}^{\ell}\, \hat{b}_{\ell m}^{*}\,\hat{a}_{\ell m}\,,
\end{equation}
We note that $\hat{a}_{\ell m}$ is computed using Eq. (\ref{eq:gwalm}). The corresponding cosmic variance is~\cite{Louis:2019tlz}
\begin{equation}\label{eq:cosmicvar12}
    \sigma^2_{\text{C}}\big(\hat{C}_\ell^{\text{GW-gal}}\big)=\frac{\hat{C}_\ell^{\text{GW-GW}}\, \hat{C}_\ell^{\text{gal-gal}}+\big(\hat{C}_\ell^{\text{GW-gal}}\big)^2}{(2\ell+1)\,f_{\text{sky}}}\,.
\end{equation}
Similarly to the auto-correlation case, 
the variance due to the local merger rate uncertainty, $\sigma^2_{\text{R}}\big(\hat{C}_\ell^{\text{GW-gal}}\big)$, is estimated 
from the variations in $\hat{C}_\ell^{\rm GW-gal}$ obtained by varying the merger rates of different types of binaries (BNS, BBH, BHNS) under the assumption of a Gaussian distribution.
%by the difference between the 95\% upper bound and 5\% lower bound of $\hat{C}_\ell^{\text{GW-gal}}$ calculated with the GW energy density of different types of binaries (BNS, BBH, BHNS) using the corresponding 95\% upper bound and 5\% lower bound of the local merger rate in Eq. (\ref{eq:localrate}), converting to 1 $\sigma$ by assuming a Gaussian distribution. {\color{red}{Kate: remove equation?}} 
%\begin{equation}
%    \sigma^2_{\text{R}}\big(\hat{C}_\ell^{\text{GW-gal}}\big)= \bigg(\frac{\hat{C}_\ell^{\text{up}}-\hat{C}_\ell^{\text{low}}}{2\cdot1.6}\bigg)^2\,,\;
%\end{equation}
%where $\hat{C}_\ell^{\text{up}}$ and $\hat{C}_\ell^{\text{low}}$ are computed by Eq. (\ref{eq:crosscl}) using the $\hat{a}_{\ell m}$ from the upper and lower bound GW maps treated the same way as in Eq. (\ref{eq:sigmaRClGW}), with the same $\hat{b}_{\ell m}$ as in Eq. (\ref{eq:crosscl}). 
The total variance is then
\begin{equation}\label{eq:totsigmacross}
    \sigma^2_{\text{tot}}\big(\hat{C}_\ell^{\text{GW-gal}}\big)=\sigma^2_{\text{C}}\big(\hat{C}_\ell^{\text{GW-gal}}\big)+\sigma^2_{\text{R}}\big(\hat{C}_\ell^{\text{GW-gal}}\big)\,.
\end{equation}
Figure \ref{fig:crosscl} shows the results of this calculation applied to the {\it Euclid} Flagship Simulation Galaxy Catalogue data. It also depicts the uncertainty due to the uncertainty in the local merger rates (Eq. \ref{eq:localrate}) and due to the total variance (Eq. \ref{eq:totsigmacross}). Similar to the SGWB auto-correlation case in Figure \ref{fig:gwcl}, for $\ell<40$ the total variance $\sigma^2_{\text{tot}}(\hat{C}_\ell^{\text{cross}})$ is dominated by cosmic variance of similar magnitude of $\hat{C}_\ell^{\text{cross}}$ and its lower bound falls below the display threshold on a logarithmic scale due to partial sky coverage. The wiggling trend over $\ell$ is analogous to that in the SGWB auto-correlation case (see Figure \ref{fig:gwcl}). 

Below, we will compare this simulation-based prediction with the semi-analytic astrophysical models discussed in Sect. \ref{Sec:TheoryModel}. The best-fit semi-analytic model is also shown in Figure \ref{fig:crosscl}, showing good agreement with our result for $\ell\ge 40$. At lower multipoles, the trend is not captured, but this is likely due to the partial sky coverage of the simulations, reflected in the  large  uncertainties  at $\ell<40$. The current upper limit on $\hat{C}_\ell^{\text{GW-gal}}$ set by the LIGO O3 data in the 60--70 Hz frequency band is  $\approx 10^{-9}$ for $\ell_{\text{max}}=5$ \cite[Fig.~7 of ][]{Yang:2023eqi}. This value is about $10^7$ times larger than the predicted signal, as expected by the limited sensitivity of the O3 LIGO data \cite{aLIGO:2020wna}.
%{\color{red} %CS: PLEASE CHECK the previous sentence. It would also be good to have here a quick reference to expected sensitivities in the following runs, as well as the next generation detectors.}\textcolor{blue}{VM: I think the sentence is good. Making projected sensitivities may take too long at this point, given Kate's time-scale - I suggest we skip this now...}

%\section{Parameter Estimation}\label{sec:PE}

We proceed to relate this Catalogue-based estimate with the semi-analytic model discussed in Sect. \ref{Sec:TheoryModel}. The semi-analytic model depends on the astrophysical kernel $\mathcal{A}(f,z)$ and the window function $W(z)$. The window function is a flat window with an observational redshift limit cut, which in our case is $z=3$. 

In \cite{Cusin:2019jpv} different astrophysical models from binary compact objects formation and evolution have been explored. Each model was derived from a reference model by varying one key aspect, while keeping the others fixed, and adjusting an efficiency factor such that all of the models result in the same total number of detectable events. This exploratory approach identified a set of parameters as key in determining the angular power spectrum of the anisotropies. In particular, common features in the redshift dependence of the kernel (to the first approximation) can be captured by the following Gaussian parameterization
\begin{equation}\label{eq:A_z_f}
	    \mathcal{A}(f,z)=\mathcal{A}(f) \,{\mathrm{e}}^{-(z-z_c)^2/2\sigma_z^2}=A_0 \, f^{-1/3} \, {\mathrm{e}}^{-(z-z_c)^2/2\sigma_z^2}\,,
\end{equation}
where we used $z=z(r)$ to express the astrophysical kernel as a function of redshift and frequency. This parametrization of the kernel includes three free parameters in total $\theta = (A_0,\,z_c,\,\sigma_z)$: kernel amplitude $A_0$, peak redshift $z_c$, and peak width $\sigma_z$. For the set of models of \cite{Cusin:2019jpv} the range of parameters of the Gaussian fit for $0<z<2$ at 63 Hz are: $A_0\in[1,3]\times10^{-37}\text{erg\,cm}^{-3}\,\text{s}^{-1/3},\,z_c\in[0.3,1.8],\,\sigma_z\in[0.3,1.2]$.

To explore this parameter space in comparison with our Catalogue-based results, we define the logarithmic likelihood following the approach of \cite{Yang:2023eqi}. To simplify the notation, we omit the subscript ``GW-gal", understanding that all power spectra are the ones of the cross-correlation. We denote the angular power spectrum predicted from the Catalogue as $\hat{C}_\ell$, and we consider a perfect experiment where the covariance matrix is diagonal and given by the total variance as in Eq. (\ref{eq:totsigmacross}). Since the covariance is diagonal, the logarithmic likelihood can be rewritten as
\begin{equation}
    \ln\mathcal{L}=-\frac{1}{2}\ln{\lvert K_C\rvert}-\frac{1}{2}(C_\ell^{t}-\hat{C}_\ell)^{T}\,K_C^{-1}\,(C_\ell^{t}-\hat{C}_\ell)
    =-\frac{1}{2}\sum_{\ell} \left[ {\ln{\sigma^2(\hat{C}_\ell)}}+\frac{(C_\ell^{t}-\hat{C}_\ell)^2}{\sigma^2(\hat{C}_\ell)}\;\right]\,,
\end{equation}
where the $C_\ell^{t}$ is the theoretical angular power spectrum of the cross-correlation \cite{Cusin:2019jpv, Yang:2023eqi}, dependent on the three parameters $(A_0,\,z_c,\,\sigma_z)$ of the Gaussian astrophysical kernel. 

\begin{figure}[!ht]
    \centering
    \includegraphics[width=0.9\linewidth]{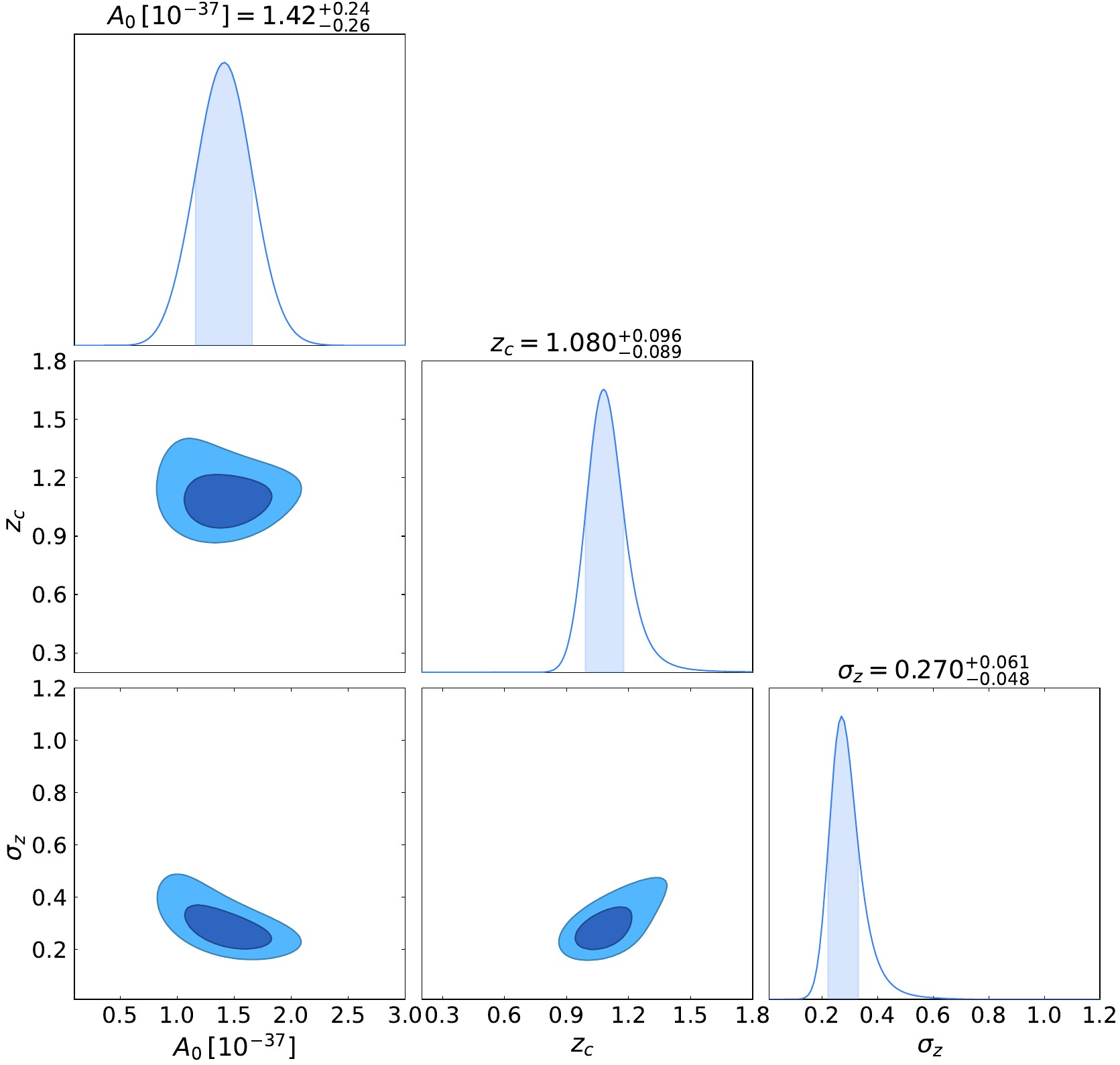}
    \caption{Contour plots using ChainConsumer~\cite{Hinton2016} in 3D parameter space of $(A_0,\,z_c,\,\sigma_z)$ for GW fluctuation $\times$ galaxy over-density cross-correlation $\hat{C}_\ell$ and astrophysical model $C_\ell(\theta),\;\theta=(A_0,\,z_c,\,\sigma_z)$ with uniform priors for all parameters.}
    \label{fig:PE}
\end{figure}

We then scan the 3-dimensional parameter space $(A_0,\,z_c,\,\sigma_z)$ with uniform priors of all three parameters with $A_0\in [1\times10^{-38},5\times10^{-37}]\,\text{erg\,cm}^{-3}\,\text{s}^{-1/3}, z_c\in[0.20,1.80], \sigma_z\in[0.01,1.20]$. The resulting posterior probability curves and contour plots are presented in Fig. \ref{fig:PE}. The best-fit parameters are $A_0=1.4\times10^{-37}\text{erg\,cm}^{-3}\,\text{s}^{-1/3},\,z_c=1.08,\,\sigma_z=0.27$. These results have to be understood as best-fit parameters for a perfect experiment observing the {\it Euclid} Flagship Simulation Galaxy Catalogue of galaxies, with zero instrumental noise and cosmic-variance limited. The model $C_\ell^{\mathrm{t}}(\theta)$ with the best-fit parameters are also shown in Fig. \ref{fig:crosscl} in red dashed line, we can see that it matches our cross-correlation estimator $\hat{C}_\ell^{\mathrm{GW-gal}}$ for $\ell$ larger than 40. We observe that the best-fit value for the amplitude $A_0$ lies within the range of values spanned by the astrophysical models explored in \cite{Cusin:2019jpv}. 

\section{Discussion and Conclusion}\label{sec:discussion}

The origin of compact binaries remains uncertain and is a subject of extensive debate. Broadly, two primary formation channels can be identified: the isolated channel, where compact binaries emerge as a byproduct of binary stellar evolution, and the dynamical channel, in which compact binaries—or their progenitors—form through gravitational interactions in dense stellar environments. %Star clusters, with significantly higher densities than those found in galactic fields, act as ideal factories for the production of dynamical CBCs. 
These differing formation processes are expected to be reflected in the source distribution and hence in the SGWB spectrum. 

Cross-correlation techniques can provide valuable insights into the formation channels of binary systems. Firstly, they enable tomographic redshift reconstruction. Furthermore, by selecting specific subpopulations from a catalogue of galaxies, this method can help identify which groups are responsible for the majority of gravitational wave emissions.

%In this paper we predicted the LVK detection of GW energy density from the sources in \textit{Euclid} simulated catalogue using the model in \cite{Jenkins:2018uac,KAGRA:2021kbb} and get a new model of 
In this paper, we estimated the amplitude and spatial anisotropy in the SGWB energy density due to compact binary coalescence events, making use of the {\textit{Euclid}} Flagship Simulation Galaxy Catalogue (version 2.1.10) \cite{Euclid:2024few}. 
%For each galaxy in the catalogue, we used the simulated mass and star-formation to constrain the galaxy's SFH and predict its contribution to the SGWB. 
We implemented an SFH model $\psi(z)$ for each galaxy, driven by its current star-formation rate and total stellar mass specified in the Catalogue. We combined the SFH model with a distribution of time delay $p(t_{\text{d}})$ between binary formation and merger, and with the distribution of metallicity $p(\mathcal{Z})$ to obtain the merger rate of compact binaries hosted by the galaxies. We then used the latest estimates of the mass distribution for compact binaries, provided by Advanced LIGO and Advanced Virgo observations through GWTC-3 \cite{KAGRA:2021duu}, and followed the method of \cite{Jenkins:2018kxc} to compute the SGWB energy density due to the entire population of merging compact binaries. Using this method, we predicted both the angular power spectrum of the resulting SGWB anisotropy and its cross-correlation with the galaxy distribution. 
%These angular spectra contain astrophysical information defined by the SFH $\psi$, delayed time distribution $p(t_{\text{d}})$, and metallicity distribution $p(\mathcal{Z})$: 

Finally, we compared our predictions with previous results in the literature obtained by using the {\it Millennium} simulation, and to predictions of semi-analytic models such as the ones of \cite{Cusin:2019jpv}. After direct inspection, we find that the astrophysical dependence can be effectively parametrised introducing a Gaussian kernel, dependent on three parameters. Assuming a perfect experiment with zero instrumental noise, we applied a Bayesian approach to explore the parameter space of this kernel and set constraints on the model parameters. The resulting best-fit values are within the realistic range of values predicted in the explorative approach of \cite{Cusin:2019jpv}. 

This exercise serves as an initial step toward developing a consistent framework for utilizing future {\it Euclid} data. We showed that future measurements of the SGWB angular power spectrum and the SGWB-galaxy correlation angular power spectrum can be used to constrain the combination of $\psi$, $p(t_{\text{d}})$, and $p(\mathcal{Z})$. Moreover, combining these measurements with other observations (e.g. SFH measurements by {\it Euclid}) can then further constrain the remaining quantities. 
The next step will involve incorporating simulated noise components into the {\it Euclid} Flagship Simulation Galaxy Catalogue, and replicating realistic LVK background noise. This preparation will enhance our ability to analyze both galaxy and background data, allowing us to anticipate the expected outcomes of future correlations.

\section*{Acknowledgments}
%\AckCosmoHub
This work has made use of CosmoHub, developed by PIC (maintained by IFAE and CIEMAT) in collaboration with ICE-CSIC. CosmoHub received funding from the Spanish government (MCIN/AEI/10.13039/501100011033), the EU NextGeneration/PRTR (PRTR-C17.I1), and the Generalitat de Catalunya.
%This work has made use of CosmoHub.
%CosmoHub has been developed by the Port d'Informació Científica (PIC), maintained through a collaboration of the Institut de Física d'Altes Energies (IFAE) and the Centro de Investigaciones Energéticas, Medioambientales y Tecnológicas (CIEMAT) and the Institute of Space Sciences (CSIC and IEEC).
%CosmoHub was partially funded by the "Plan Estatal de Investigación Científica y Técnica y de Innovación" program of the Spanish government, has been supported by the call for grants for Scientific and Technical Equipment 2021 of the State Program for Knowledge Generation and Scientific and Technological Strengthening of the R+D+i System, financed by MCIN/AEI/ 10.13039/501100011033 and the EU NextGeneration/PRTR (Hadoop Cluster for the comprehensive management of massive scientific data, reference EQC2021-007479-P) and by MICIIN with funding from European Union NextGenerationEU(PRTR-C17.I1) and by Generalitat de Catalunya. The work of GC is supported by CNRS and by the SNSF Ambizione grant \emph{Gravitational wave propagation in the clustered Universe}. 

The authors are grateful for computational resources provided by the LIGO Laboratory (CIT) supported by National Science Foundation Grants PHY-0757058 and PHY-0823459 and Inter-University Center for Astronomy and Astrophysics (Sarathi).%LIGO clusters (Caltech)

The work of KY, CS, and VM was in part supported by NSF grants PHY-2308486 and PHY-2011675.

The Euclid Consortium acknowledges the European Space Agency and a number of agencies and institutes that have supported the development of {\it Euclid}, in particular the Agenzia Spaziale Italiana, the Austrian Forschungsf\"orderungsgesellschaft funded through BMIMI, the Belgian Science Policy, the Canadian Euclid Consortium, the Deutsches Zentrum f\"ur Luft- und Raumfahrt, the DTU Space and the Niels Bohr Institute in Denmark, the French Centre National d'Etudes Spatiales, the Funda\c{c}\~{a}o para a Ci\^{e}ncia e a Tecnologia, the Hungarian Academy of Sciences, the Ministerio de Ciencia, Innovaci\'{o}n y Universidades, the National Aeronautics and Space Administration, the National Astronomical Observatory of Japan, the Netherlandse Onderzoekschool Voor Astronomie, the Norwegian Space Agency, the Research Council of Finland, the Romanian Space Agency, the State Secretariat for Education, Research, and Innovation (SERI) at the Swiss Space Office (SSO), and the United Kingdom Space Agency. A complete and detailed list is available on the {\it Euclid} web site (\url{www.euclid-ec.org}).

\bibliographystyle{unsrt}%{abbrv}
\bibliography{duplicate}%{EuclidCBC}

@unpublished{Cusin:2025xle,
    author = "G. Cusin and C. Pitrou and M. Pijnenburg et al.",
    title = "{Measuring anisotropies in the PTA band with cross-correlations}",
    eprint = "2502.17401",
    archivePrefix = "arXiv",
    primaryClass = "gr-qc",
    month = Feb,
    year = "2025",
    note = "arXiv:2502.17401"
}

@article{Yang:2023eqi,
    author = "K. Z. Yang and J. Suresh and G. Cusin et al.",
    title = "{Measurement of the cross-correlation angular power spectrum between the stochastic gravitational wave background and galaxy overdensity}",
    eprint = "2304.07621",
    archivePrefix = "arXiv",
    primaryClass = "gr-qc",
    doi = "10.1103/PhysRevD.108.043025",
    journal = "Phys. Rev. D",
    volume = "108",
    number = "4",
    pages = "043025",
    year = "2023",
    month = aug
}

@article{Louis:2019tlz,
    author = "T. Louis and X. Garrido and A. Soussana et al.",
    title = "{Consistency of CMB experiments beyond cosmic variance}",
    eprint = "1905.06864",
    archivePrefix = "arXiv",
    primaryClass = "astro-ph.CO",
    doi = "10.1103/PhysRevD.100.023518",
    journal = "Phys. Rev. D",
    volume = "100",
    number = "2",
    pages = "023518",
    year = "2019",
    month = jul
}

@article{Alonso:2024knf,
    author = "D. Alonso and M. Nikjoo and A. I. Renzini et al.",
    title = "{Tomographic constraints on the production rate of gravitational waves from astrophysical sources}",
    eprint = "2406.19488",
    archivePrefix = "arXiv",
    primaryClass = "astro-ph.CO",
    doi = "10.1103/PhysRevD.110.103544",
    journal = "Phys. Rev. D",
    volume = "110",
    number = "10",
    pages = "103544",
    year = "2024",
    month = nov
}

@article{Pitrou:2024scp,
    author = "C. Pitrou and Cusin, G.",
    title = "{Mitigating cosmic variance in the Hellings-Downs curve: A cosmic microwave background analogy}",
    eprint = "2412.12073",
    archivePrefix = "arXiv",
    primaryClass = "gr-qc",
    doi = "10.1103/PhysRevD.111.083544",
    journal = "Phys. Rev. D",
    volume = "111",
    number = "8",
    pages = "083544",
    year = "2025",
    month = Apr
}

@article{Cusin:2024git,
    author = "G. Cusin and C. Pitrou and C. Bonvin et al.",
    title = "{Boosting gravitational waves: a review of kinematic effects on amplitude, polarization, frequency and energy density}",
    eprint = "2405.01297",
    archivePrefix = "arXiv",
    primaryClass = "gr-qc",
    doi = "10.1088/1361-6382/ad7ad0",
    journal = "Class. Quant. Grav.",
    volume = "41",
    number = "22",
    pages = "225006",
    year = "2024",
    month = oct
}

@article{Tallada:2020qmg,
    author = "P. Tallada and J. Carretero and J. Casals et al.",
    title = "{CosmoHub: Interactive exploration and distribution of astronomical data on Hadoop}",
    eprint = "2003.03217",
    archivePrefix = "arXiv",
    primaryClass = "astro-ph.IM",
    doi = "10.1016/j.ascom.2020.100391",
    journal = "Astron. Comput.",
    volume = "32",
    pages = "100391",
    year = "2020",
    month = jul
}

@article{Carretero:2017zkw,
    author = "J. Carretero and P. Tallada and J. Casals et al.",
    title = "{CosmoHub and SciPIC: Massive cosmological data analysis, distribution and generation using a Big Data platform}",
    doi = "10.22323/1.314.0488",
    journal = "PoS",
    volume = "EPS-HEP2017",
    pages = "488",
    year = "2017",
    month = Jul
}

@article{Euclid:2024few,
    author = "Euclid Collaboration: F. J. Castander and P. Fosalba and J. Stadel et al.",
    collaboration = "Euclid",
    title = "{Euclid: V. The Flagship galaxy mock catalogue: A comprehensive simulation for the Euclid mission}",
    journal = {A\&A},
    year = 2025,
    month = may,
    volume = {697},
    eid = {A5},
    pages = {A5},
    doi = {10.1051/0004-6361/202450853},
    archivePrefix = {arXiv},
    eprint = {2405.13495},
    primaryClass = {astro-ph.CO},
}

@ARTICLE{Hinton2016,
   author = {S.~R.~{Hinton}},
    title = "{ChainConsumer}",
  journal = {The Journal of Open Source Software},
     year = 2016,
    month = Aug,
   volume = 1,
      eid = {00045},
    pages = {00045},
      doi = {10.21105/joss.00045},
}

@article{Yang:2020usq,
    author = "K. Z. Yang and V. Mandic and C. Scarlata et al.",
    title = "{Searching for Cross-Correlation Between Stochastic Gravitational Wave Background and Galaxy Number Counts}",
    eprint = "2007.10456",
    archivePrefix = "arXiv",
    primaryClass = "astro-ph.CO",
    doi = "10.1093/mnras/staa3159",
    journal = "MNRAS",
    volume = "500",
    number = "2",
    pages = "1666--1672",
    year = 2021,
    month = jan,
}

@article{Allende_Prieto_2001,
    author = {C. Allende Prieto and D. L. Lambert and M. Asplund},
    title = {The Forbidden Abundance of Oxygen in the Sun},
    doi = {10.1086/322874},
    url = {https://dx.doi.org/10.1086/322874},
    journal = {ApJL},
    year = {2001},
    month = Jul,
    volume = {556},
    number = {1},
    pages = {L63--L66},
}

@ARTICLE{Curti2020,
    author = {M. {Curti} and F. {Mannucci} and G. {Cresci} et al.},
    title = "{The mass-metallicity and the fundamental metallicity relation revisited on a fully T$_{e}$-based abundance scale for galaxies}",
    journal = {MNRAS},
    year = 2020,
    month = Jan,
    volume = {491},
    number = {1},
    pages = {944--964},
    doi = {10.1093/mnras/stz2910},
    archivePrefix = {arXiv},
    eprint = {1910.00597},
    primaryClass = {astro-ph.GA},
}

@ARTICLE{Thomas_2005,
       author = {D. {Thomas} and C. {Maraston} and R. {Bender} et al.},
        title = "{The Epochs of Early-Type Galaxy Formation as a Function of Environment}",
      journal = {ApJ},
         year = 2005,
        month = Mar,
       volume = {621},
       number = {2},
        pages = {673--694},
          doi = {10.1086/426932},
archivePrefix = {arXiv},
       eprint = {astro-ph/0410209},
 primaryClass = {astro-ph},
}

@article{Mason_2015,
    author = {C. A. Mason and M. Trenti and T. Treu},
    title = {THE GALAXY UV LUMINOSITY FUNCTION BEFORE THE EPOCH OF REIONIZATION},
    journal = {ApJ},
    doi = {10.1088/0004-637X/813/1/21},
    url = {https://dx.doi.org/10.1088/0004-637X/813/1/21},
    year = {2015},
    month = nov,
    publisher = {The American Astronomical Society},
    volume = {813},
    number = {1},
    pages = {21}
}

@ARTICLE{Vangioni:2014axa,
       author = {E. {Vangioni} and K. A. {Olive} and T. {Prestegard} et al.},
        title = "{The impact of star formation and gamma-ray burst rates at high redshift on cosmic chemical evolution and reionization}",
      journal = {MNRAS},
      year = 2015,
        month = Mar,
       volume = {447},
       number = {3},
        pages = {2575--2587},
          doi = {10.1093/mnras/stu2600},
archivePrefix = {arXiv},
       eprint = {1409.2462},
 primaryClass = {astro-ph.GA},
}

@ARTICLE{Madau2014,
       author = {P. {Madau} and M. {Dickinson}},
        title = "{Cosmic Star-Formation History}",
      journal = {ARA\&A},
         year = 2014,
        month = Aug,
       volume = {52},
        pages = {415--486},
          doi = {10.1146/annurev-astro-081811-125615},
archivePrefix = {arXiv},
       eprint = {1403.0007},
 primaryClass = {astro-ph.CO},
}

@article{Renzini:2006je,
    author = "A. Renzini",
    title = "{Stellar population diagnostics of elliptical galaxy formation}",
    eprint = "astro-ph/0603479",
    archivePrefix = "arXiv",
    doi = "10.1146/annurev.astro.44.051905.092450",
    journal = {ARA\&A},
    volume = "44",
    number = {1},
    pages = {141--192},
    year = "2006",
    month = sep
}

@article{Diemer:2017ttl,
    author = "B. Diemer and M. Sparre and L. E. Abramson et al.",
    title = "{Log-normal star formation histories in simulated and observed galaxies}",
    eprint = "1701.02308",
    archivePrefix = "arXiv",
    primaryClass = "astro-ph.GA",
    doi = "10.3847/1538-4357/aa68e5",
    journal = "ApJ",
    volume = "839",
    number = "1",
    pages = "26",
    year = "2017",
    month = apr
}

@article{aLIGO:2020wna,
    author = "A. Buikema and C. Cahillane and G. L. Mansell and others",
    collaboration = "aLIGO",
    title = "{Sensitivity and performance of the Advanced LIGO detectors in the third observing run}",
    eprint = "2008.01301",
    archivePrefix = "arXiv",
    primaryClass = "astro-ph.IM",
    doi = "10.1103/PhysRevD.102.062003",
    journal = "Phys. Rev. D",
    volume = "102",
    number = "6",
    pages = "062003",
    year = "2020",
    month = sep
}

@article{KAGRA:2021mth,
    author = "R. Abbott and T.~D. {Abbott} and S. {Abraham} et al.",
    collaboration = "KAGRA, Virgo, LIGO Scientific",
    title = "{Search for anisotropic gravitational-wave backgrounds using data from Advanced LIGO and Advanced Virgo\textquoteright{}s first three observing runs}",
    eprint = "2103.08520",
    archivePrefix = "arXiv",
    primaryClass = "gr-qc",
    reportNumber = "LIGO-P2000500",
    doi = "10.1103/PhysRevD.104.022005",
    journal = "Phys. Rev. D",
    volume = "104",
    number = "2",
    pages = "022005",
    year = "2021",
    month = jul
}

@article{KAGRA:2021duu,
    author = "R. Abbott and T.~D. Abbott and F. Acernese et al.",
    collaboration = "KAGRA, VIRGO, LIGO Scientific",
    title = "{Population of Merging Compact Binaries Inferred Using Gravitational Waves through GWTC-3}",
    eprint = "2111.03634",
    archivePrefix = "arXiv",
    primaryClass = "astro-ph.HE",
    reportNumber = "LIGO-P2100239 ; Data release: https://zenodo.org/record/5655785, LIGO-P2100239",
    doi = "10.1103/PhysRevX.13.011048",
    journal = "Phys. Rev. X",
    volume = "13",
    number = "1",
    pages = "011048",
    year = "2023",
    month = jan
}

@article{KAGRA:2020agh,
    author = "T. Akutsu and M. Ando and K. Arai et al.",
    collaboration = "KAGRA",
    title = "{Overview of {KAGRA}: Calibration, detector characterization, physical environmental monitors, and the geophysics interferometer}",
    eprint = "2009.09305",
    archivePrefix = "arXiv",
    primaryClass = "gr-qc",
    doi = "10.1093/ptep/ptab018",
    journal = "Progress of Theoretical and Experimental Physics",
    volume = "2021",
    number = "5",
    pages = "05A102",
    year = "2021",
    month = may
}

@article{Alonso:2020mva,
    author = "D. Alonso and G. Cusin and P. G. Ferreira et al.",
    title = "{Detecting the anisotropic astrophysical gravitational wave background in the presence of shot noise through cross-correlations}",
    eprint = "2002.02888",
    archivePrefix = "arXiv",
    primaryClass = "astro-ph.CO",
    doi = "10.1103/PhysRevD.102.023002",
    journal = "Phys. Rev. D",
    volume = "102",
    number = "2",
    pages = "023002",
    year = "2020",
    month = jul
}

@article{Capurri:2021zli,
    author = "G. Capurri and A. Lapi and C. Baccigalupi et al.",
    title = "{Intensity and anisotropies of the stochastic gravitational wave background from merging compact binaries in galaxies}",
    eprint = "2103.12037",
    archivePrefix = "arXiv",
    primaryClass = "gr-qc",
    doi = "10.1088/1475-7516/2021/11/032",
    journal = "JCAP",
    volume = "11",
    pages = "032",
    year = "2021",
    month = nov
}

@article{Sah:2023bgr,
    author = "M. R. Sah and S. Mukherjee",
    title = "{Non-stationary astrophysical stochastic gravitational-wave background: a new probe to the high-redshift population of binary black holes}",
    journal = {MNRAS},
    year = 2024,
    month = jan,
    volume = {527},
    number = {2},
    pages = {4100--4111},
    doi = {10.1093/mnras/stad3365},
    archivePrefix = {arXiv},
    eprint = {2307.06405},
    primaryClass = {gr-qc},
}

@article{Mukherjee:2019oma,
    author = "S. Mukherjee and J. Silk",
    title = "{Time-dependence of the astrophysical stochastic gravitational wave background}",
    eprint = "1912.07657",
    archivePrefix = "arXiv",
    primaryClass = "gr-qc",
    doi = "10.1093/mnras/stz3226",
    journal = "MNRAS",
    volume = "491",
    number = "4",
    pages = "4690--4701",
    year = "2020",
    month = nov
}

@article{WiggleZ:2013kor,
    author = "F. A. Marin and C. Blake and G. B. Poole et al.",
    collaboration = "WiggleZ",
    title = "{The WiggleZ Dark Energy Survey: constraining galaxy bias and cosmic growth with 3-point correlation functions}",
    eprint = "1303.6644",
    archivePrefix = "arXiv",
    primaryClass = "astro-ph.CO",
    doi = "10.1093/mnras/stt520",
    journal = "MNRAS",
    volume = "432",
    number = {4},
    pages = {2654--2668},
    year = "2013",
    month = jul
}

@unpublished{Rassat:2008ja,
    author = "A. Rassat and A. Amara and L. Amendola et al.",
    title = "{Deconstructing Baryon Acoustic Oscillations: A Comparison of Methods}",
    eprint = "0810.0003",
    archivePrefix = "arXiv",
    primaryClass = "astro-ph",
    month = Oct,
    year = "2008",
    note = "arXiv:0810.0003"
}

@article{Jenkins:2018kxc,
      author         = "A. C. Jenkins and R. O'Shaughnessy and M. Sakellariadou et al.",
      title          = "{Anisotropies in the astrophysical gravitational-wave
                        background: The impact of black hole distributions}",
      journal        = "Phys. Rev. Lett.",
      volume         = "122",
      year           = "2019",
      month          = mar,
      number         = "11",
      pages          = "111101",
      doi            = "10.1103/PhysRevLett.122.111101",
      eprint         = "1810.13435",
      archivePrefix  = "arXiv",
      primaryClass   = "astro-ph.CO",
      reportNumber   = "KCL-PH-TH/2018-60",
      SLACcitation   = "%%CITATION = ARXIV:1810.13435;%%"
}

@article{Jenkins:2018uac,
      author         = "A. C. Jenkins and M. Sakellariadou and T. Regimbau et al.",
      title          = "{Anisotropies in the astrophysical gravitational-wave
                        background: Predictions for the detection of compact
                        binaries by {LIGO} and {Virgo}}",
      journal        = "Phys. Rev.",
      volume         = "D98",
      year           = "2018",
      month          = sep,
      number         = "6",
      pages          = "063501",
      doi            = "10.1103/PhysRevD.98.063501",
      eprint         = "1806.01718",
      archivePrefix  = "arXiv",
      primaryClass   = "astro-ph.CO",
      reportNumber   = "KCL-PH-TH/2018-23, KCL-PH-TH-2018-23",
      SLACcitation   = "%%CITATION = ARXIV:1806.01718;%%"
}

@article{GW170817implySGWB,
    title = {GW170817: Implications for the Stochastic Gravitational-Wave Background from Compact Binary Coalescences},
    author = {B. P. Abbott and R. Abbott and T. D. Abbott and others},
    collaboration = {LIGO Scientific Collaboration and Virgo Collaboration},
    journal = {Phys. Rev. Lett.},
    volume = {120},
    number = {9},
    pages = {091101},
    numpages = {12},
    year = {2018},
    month = mar,
    publisher = {American Physical Society},
    doi = {10.1103/PhysRevLett.120.091101},
    url = {https://link.aps.org/doi/10.1103/PhysRevLett.120.091101}
}

@article{O1rateUL,
    author = {B. P. Abbott and R. Abbott and T. D. Abbott and others},
    collaboration={LIGO Scientific Collaboration and Virgo Collaboration},
    title = {Upper Limits on the Rates of Binary Neutron Star and Neutron Star-Black Hole Mergers from Advanced LIGO’s First Observing Run},
    journal = {ApJL},
    doi = {10.3847/2041-8205/832/2/L21},
    url = {https://doi.org/10.3847/2041-8205/832/2/L21},
    year = {2016},
    month = dec,
    publisher = {The American Astronomical Society},
    volume = {832},
    number = {2},
    pages = {L21},
    archivePrefix = {arXiv},
    eprint = {1607.07456},
    primaryClass = {astro-ph.HE}
}

@article{Jenkins:2019nks,
      author         = "A. C. Jenkins and J. D. Romano and and M. Sakellariadou",
      title          = "{Estimating the angular power spectrum of the
                        gravitational-wave background in the presence of shot
                        noise}",
      journal        = "Phys. Rev.",
      volume         = "D100",
      year           = "2019",
      month          = oct,
      number         = "8",
      pages          = "083501",
      doi            = "10.1103/PhysRevD.100.083501",
      eprint         = "1907.06642",
      archivePrefix  = "arXiv",
      primaryClass   = "astro-ph.CO",
      reportNumber   = "KCL-PH-TH/2019-59",
}

@article{Jenkins:2019uzp,
      author         = "A. C. Jenkins and M. Sakellariadou",
      title          = "{Shot noise in the astrophysical gravitational-wave
                        background}",
      journal        = "Phys. Rev.",
      volume         = "D100",
      year           = "2019",
      month          = sep,
      number         = "6",
      pages          = "063508",
      doi            = "10.1103/PhysRevD.100.063508",
      eprint         = "1902.07719",
      archivePrefix  = "arXiv",
      primaryClass   = "astro-ph.CO",
      SLACcitation   = "%%CITATION = ARXIV:1902.07719;%%"
}

@article{Cusin:2018rsq,
      author         = "G. Cusin and I. Dvorkin and C. Pitrou et al.",
      title          = "{First predictions of the angular power spectrum of the
                        astrophysical gravitational wave background}",
      journal        = "Phys. Rev. Lett.",
      volume         = "120",
      year           = "2018",
      month          = jun,
      number         = {23},
      pages          = "231101",
      doi            = "10.1103/PhysRevLett.120.231101",
      eprint         = "1803.03236",
      archivePrefix  = "arXiv",
      primaryClass   = "astro-ph.CO",
      SLACcitation   = "%%CITATION = ARXIV:1803.03236;%%"
}

@article{Cusin:2017mjm,
    author = "G. Cusin and C. Pitrou and J.-P. Uzan",
    title = "{The signal of the gravitational wave background and the angular correlation of its energy density}",
    eprint = "1711.11345",
    archivePrefix = "arXiv",
    primaryClass = "astro-ph.CO",
    doi = "10.1103/PhysRevD.97.123527",
    journal = "Phys. Rev. D",
    volume = "97",
    number = "12",
    pages = "123527",
    year = "2018",
    month = jun
}

@article{Cusin:2019jpv,
    author = "G. Cusin and I. Dvorkin and C. Pitrou et al.",
    title = "{Properties of the stochastic astrophysical gravitational wave background: astrophysical sources dependencies}",
    eprint = "1904.07797",
    archivePrefix = "arXiv",
    primaryClass = "astro-ph.CO",
    doi = "10.1103/PhysRevD.100.063004",
    journal = "Phys. Rev. D",
    volume = "100",
    number = "6",
    pages = "063004",
    year = "2019",
    month = sep
}

@article{Cusin:2018avf,
    author = "G. Cusin and R. Durrer and P. G. Ferreira",
    title = "{Polarization of a stochastic gravitational wave background through diffusion by massive structures}",
    eprint = "1807.10620",
    archivePrefix = "arXiv",
    primaryClass = "astro-ph.CO",
    doi = "10.1103/PhysRevD.99.023534",
    journal = "Phys. Rev. D",
    volume = "99",
    number = "2",
    pages = "023534",
    year = "2019",
    month = jan
}

@article{Cusin:2019jhg,
    author = "G. Cusin and I. Dvorkin and C. Pitrou et al.",
    title = "{Stochastic gravitational wave background anisotropies in the mHz band: astrophysical dependencies}",
    eprint = "1904.07757",
    archivePrefix = "arXiv",
    primaryClass = "astro-ph.CO",
    doi = "10.1093/mnrasl/slz182",
    journal = "MNRAS",
    volume = "493",
    number = "1",
    pages = "L1--L5",
    year = "2020",
    month = mar
}

@article{Cusin:2017fwz,
    author = "G. Cusin and C. Pitrou and J.-P. Uzan",
    title = "{Anisotropy of the astrophysical gravitational wave background: Analytic expression of the angular power spectrum and correlation with cosmological observations}",
    eprint = "1704.06184",
    archivePrefix = "arXiv",
    primaryClass = "astro-ph.CO",
    doi = "10.1103/PhysRevD.96.103019",
    journal = "Phys. Rev. D",
    volume = "96",
    number = "10",
    pages = "103019",
    year = "2017",
    month = nov
}

@article{Pitrou:2019rjz,
    author = "C. Pitrou and G. Cusin and Uzan, J. P.",
    title = "{Unified view of anisotropies in the astrophysical gravitational-wave background}",
    eprint = "1910.04645",
    archivePrefix = "arXiv",
    primaryClass = "astro-ph.CO",
    doi = "10.1103/PhysRevD.101.081301",
    journal = "Phys. Rev. D",
    volume = "101",
    number = "8",
    pages = "081301",
    year = "2020",
    month = apr
}

@article{Zonca2019,
    author = {A. Zonca and L. Singer and D. Lenz et al.},
    title = {healpy: equal area pixelization and spherical harmonics transforms for data on the sphere in Python},
    journal = {Journal of Open Source Software},
    doi = {10.21105/joss.01298},
    url = {https://doi.org/10.21105/joss.01298},
    year = {2019},
    month = Mar,
    publisher = {The Open Journal},
    volume = {4},
    number = {35},
    pages = {1298},
}

@article{HEALPix,
   author = {K.~M. {G{\'o}rski} and E. {Hivon} and A.~J. {Banday} et al.},
    title = "{HEALPix: A Framework for High-Resolution Discretization and Fast Analysis of Data Distributed on the Sphere}",
  journal = {ApJ},
   eprint = {arXiv:astro-ph/0409513},
     year = 2005,
    month = Apr,
   volume = 622,
    number = {2},
    pages = {759--771},
      doi = {10.1086/427976},
}

@article{aLIGO,
      author = {{LIGO Scientific Collaboration}},
        title = "{{A}dvanced {LIGO}}",
      journal = {Classical and Quantum Gravity},
         year = 2015,
        month = Apr,
      volume = {32},
      number = {7},
          eid = {074001},
        pages = {074001},
          doi = {10.1088/0264-9381/32/7/074001},
archivePrefix = {arXiv},
      eprint = {1411.4547},
 primaryClass = {gr-qc},
}

@article{avirgo,
    author = {F. Acernese and M. Agathos and K. Agatsuma et al.},
    journal = {Classical and Quantum Gravity},
    collaboration = "VIRGO",
    title = "{Advanced Virgo: a second-generation interferometric gravitational wave detector}",
    eprint = "1408.3978",
    archivePrefix = "arXiv",
    primaryClass = "gr-qc",
    doi = "10.1088/0264-9381/32/2/024001",
    volume = "32",
    number = "2",
    pages = "024001",
    year = "2015",
    month = jan
}

@unpublished{GWTC4,
    author = "The LIGO Scientific Collaboration and the Virgo Collaboration and the KAGRA Collaboration",
    title = "{GWTC-4.0: Updating the Gravitational-Wave Transient Catalog with Observations from the First Part of the Fourth LIGO-Virgo-KAGRA Observing Run}",
    eprint = "2508.18082",
    archivePrefix = "arXiv",
    primaryClass = "gr-qc",
    reportNumber = "LIGO-P2400386",
    month = aug,
    year = "2025",
    note = "arXiv:2508.18082"
}

@unpublished{GWTC4pop,
      title={GWTC-4.0: Population Properties of Merging Compact Binaries}, 
      author={The LIGO Scientific Collaboration and the Virgo Collaboration and the KAGRA Collaboration},
      year={2025},
      month = aug,
      eprint={2508.18083},
      archivePrefix={arXiv},
      primaryClass={astro-ph.HE},
      url={https://arxiv.org/abs/2508.18083}, 
      note = "arXiv:2508.18083"
}

@unpublished{GWTC4H0,
      title={GWTC-4.0: Constraints on the Cosmic Expansion Rate and Modified Gravitational-wave Propagation}, 
      author={The LIGO Scientific Collaboration and the Virgo Collaboration and the KAGRA Collaboration},
      year={2025},
      month = sep,
      eprint={2509.04348},
      archivePrefix={arXiv},
      primaryClass={astro-ph.CO},
      url={https://arxiv.org/abs/2509.04348},
      note = "arXiv:2509.04348"
}

@unpublished{O4a-isoback,
    author = "The LIGO Scientific Collaboration and the Virgo Collaboration and the KAGRA Collaboration",
    title = "{Upper Limits on the Isotropic Gravitational-Wave Background from the first part of LIGO, Virgo, and KAGRA's fourth Observing Run}",
    journal = {arXiv e-prints},
    eprint = "2508.20721",
    archivePrefix = "arXiv",
    primaryClass = "gr-qc",
    reportNumber = "LIGO-P2500349",
    month = aug,
    year = "2025",
    note = "arXiv:2508.20721"
}

@unpublished{O3TGR,
    author = "{The LIGO Scientific Collaboration} and {the Virgo Collaboration} and {the KAGRA Collaboration}",
    title = "{Tests of General Relativity with GWTC-3}",
    eprint = "2112.06861",
    archivePrefix = "arXiv",
    primaryClass = "gr-qc",
    reportNumber = "LIGO-P2100275",
    month = Dec,
    year = "2021",
    note = "arXiv:2112.06861"
}

@article{GW170817_EOS,
  title = {{GW170817}: Measurements of Neutron Star Radii and Equation of State},
  author = {B. P. Abbott and R. Abbott and T.~D. Abbott et al.},
  collaboration = {The LIGO Scientific Collaboration and the Virgo Collaboration},
  journal = {Phys. Rev. Lett.},
  volume = {121},
  number = {16},
  pages = {161101},
  numpages = {16},
  year = {2018},
  month = Oct,
  publisher = {American Physical Society},
  doi = {10.1103/PhysRevLett.121.161101},
  url = {https://link.aps.org/doi/10.1103/PhysRevLett.121.161101}
}

@article{regfrei,
	author = {T. Regimbau and J. A. de Freitas Pacheco},
	title = {Stochastic Background from Coalescences of Neutron Star{\textendash}Neutron Star Binaries},
	journal = {ApJ},
    doi = {10.1086/500190},
	url = {https://doi.org/10.1086/500190},
	year = 2006,
	month = May,
	publisher = {American Astronomical Society},
	volume = {642},
	number = {1},
	pages = {455--461}
}

@article{zhu_cbc,
	author = {X.-J. Zhu and E. Howell and T. Regimbau et al.},
	title = {Stochastic Gravitational Wave Background from Coalescing Binary Black Holes},
	journal = {ApJ},
	doi = {10.1088/0004-637x/739/2/86},
	url = {https://doi.org/10.1088/0004-637x/739/2/86},
	year = 2011,
	month = oct,
	publisher = {American Astronomical Society},
	volume = {739},
	number = {2},
	pages = {86},
    archivePrefix = {arXiv},
    eprint = {1104.3565},
    primaryClass = {gr-qc}
}

@article{marassi_cbc,
  title = {Imprint of the merger and ring-down on the gravitational wave background from black hole binaries coalescence},
  author = {S. Marassi and R. Schneider and G. Corvino et al.},
  journal = {Phys. Rev. D},
  volume = {84},
  number = {12},
  pages = {124037},
  numpages = {14},
  year = {2011},
  month = Dec,
  publisher = {American Physical Society},
  doi = {10.1103/PhysRevD.84.124037},
  url = {https://link.aps.org/doi/10.1103/PhysRevD.84.124037},
  archivePrefix = {arXiv},
  eprint = {1111.6125},
  primaryClass = {astro-ph.CO}
}

@article{wu_cbc,
  title = {Accessibility of the gravitational-wave background due to binary coalescences to second and third generation gravitational-wave detectors},
  author = {C. Wu and V. Mandic and T. Regimbau},
  journal = {Phys. Rev. D},
  volume = {85},
  number = {10},
  pages = {104024},
  numpages = {8},
  year = {2012},
  month = May,
  publisher = {American Physical Society},
  doi = {10.1103/PhysRevD.85.104024},
  url = {https://link.aps.org/doi/10.1103/PhysRevD.85.104024},
  archivePrefix = {arXiv},
  eprint = {1112.1898},
  primaryClass = {gr-qc},
}

@article{cutler,
  title = {Gravitational waves from neutron stars with large toroidal $B$ fields},
  author = {C. Cutler},
  journal = {Phys. Rev. D},
  volume = {66},
  number = {8},
  pages = {084025},
  numpages = {6},
  year = {2002},
  month = Oct,
  publisher = {American Physical Society},
  doi = {10.1103/PhysRevD.66.084025},
  url = {https://link.aps.org/doi/10.1103/PhysRevD.66.084025},
  archivePrefix = {arXiv},
  eprint = {gr-qc/0206051},
  primaryClass = {gr-qc}
}

@article{bonazzola,
    title = {Gravitational waves from pulsars: emission by the magnetic field induced distortion},
    author = {S. Bonazzola and E. Gourgoulhon},
    journal = {A\&A},
    eprint = "astro-ph/9602107",
    archivePrefix = "arXiv",
    primaryClass = {astro-ph},
    doi = {10.48550/arXiv.astro-ph/9602107},
    volume = "312",
    pages = "675--690",
    year = "1996",
    month = aug,
}

@article{marassi_magnetar,
    author = {S. Marassi and R. Ciolfi and R. Schneider et al.},
    title = "{Stochastic background of gravitational waves emitted by magnetars}",
    eprint = "1009.1240",
    archivePrefix = "arXiv",
    primaryClass = "astro-ph.CO",
    doi = "10.1111/j.1365-2966.2010.17861.x",
    journal = "MNRAS",
    volume = "411",
    number = {4},
    pages = {2549--2557},
    year = "2011",
    month = mar
}

@article{maggiore,
    title = {Gravitational wave experiments and early universe cosmology},
    author = {M. Maggiore},
    journal = {Physics Reports},
    volume = {331},
    number = {6},
    pages = {283--367},
    year = {2000},
    month = jul,
    issn = {0370-1573},
    doi = {10.1016/S0370-1573(99)00102-7},
    url = {https://www.sciencedirect.com/science/article/pii/S0370157399001027},
    archivePrefix = {arXiv},
    eprint = {gr-qc/9909001},
    primaryClass = {gr-qc},
}

@article{regimbau_review,
	author = {T. Regimbau},
	title = {The astrophysical gravitational wave stochastic background},
	journal = {Res. Astron. Astrophys.},
	doi = {10.1088/1674-4527/11/4/001},
	url = {https://doi.org/10.1088/1674-4527/11/4/001},
	year = 2011,
	month = Mar,
	publisher = {{IOP} Publishing},
	volume = {11},
	number = {4},
	pages = {369--390},
    eprint = "1101.2762",
    archivePrefix = "arXiv",
    primaryClass = "astro-ph.CO",
}

@article{SNe,
    author = "D. M. Coward and R. R. Burman and D. G. Blair",
    title = "{Simulating a stochastic background of gravitational waves from neutron star formation at cosmological distances}",
    doi = "10.1046/j.1365-8711.2002.04981.x",
    journal = "MNRAS",
    volume = "329",
    number = {2},
    pages = "411--416",
    year = "2002",
    month = jan,
    issn = {0035-8711},
    url = {https://doi.org/10.1046/j.1365-8711.2002.04981.x},
    eprint = {https://academic.oup.com/mnras/article-pdf/329/2/411/18647252/329-2-411.pdf}
}

@article{grishchuk,
   author = {L.~P. Grishchuk},
    title = "{Amplification of gravitational waves in an isotropic universe}",
  journal = {Soviet Journal of Experimental and Theoretical Physics},
     year = 1975,
    month = Sep,
   volume = 40,
    pages = {409},
   adsurl = {https://ui.adsabs.harvard.edu/abs/1975JETP...40..409G}
}

@article{barkana,
   author = {R. Bar-Kana},
    title = "{Limits on direct detection of gravitational waves}",
  journal = {Phys. Rev. D},
     year = 1994,
    month = Jul,
   volume = 50,
   number = {2},
    pages = {1157--1160},
      doi = {10.1103/PhysRevD.50.1157},
archivePrefix = {arXiv},
       eprint = {astro-ph/9401050},
 primaryClass = {astro-ph},
   adsurl = {https://ui.adsabs.harvard.edu/abs/1994PhRvD..50.1157B}
}

@article{starob,
   author = {A.~A.~{Starobinski{\v i}}},
    title = "{Spectrum of relict gravitational radiation and the early state of the universe}",
  journal = {Soviet Journal of Experimental and Theoretical Physics Letters},
     year = 1979,
    month = Dec,
   volume = 30,
    pages = {682},
   adsurl = {https://ui.adsabs.harvard.edu/abs/1979JETPL..30..682S}
}

@article{turner,
    author = "M. S. Turner",
    title = "{Detectability of inflation produced gravitational waves}",
    eprint = "astro-ph/9607066",
    archivePrefix = "arXiv",
    primaryClass = {astro-ph},
    reportNumber = "FERMILAB-PUB-96-169-A, FERMILAB-PUB-96-167-A",
    doi = "10.1103/PhysRevD.55.R435",
    journal = "Phys. Rev. D",
    volume = "55",
    number = {2},
    pages = {R435--R439},
    year = "1997",
    month = jan
}

@article{peloso_parviol,
   author = {N. {Barnaby} and E. {Pajer} and M. {Peloso}},
    title = "{Gauge field production in axion inflation: Consequences for monodromy, non-Gaussianity in the {CMB}, and gravitational waves at interferometers}",
  journal = {Phys. Rev. D},
archivePrefix = "arXiv",
   eprint = {1110.3327},
 primaryClass = "astro-ph.CO",
     year = 2012,
    month = Jan,
   volume = 85,
   number = 2,
      eid = {023525},
    pages = {023525},
      doi = {10.1103/PhysRevD.85.023525},
   adsurl = {https://ui.adsabs.harvard.edu/abs/2012PhRvD..85b3525B}
}

@article{seto,
   author = {N. {Seto} and A. {Taruya}},
    title = "{Measuring a Parity-Violation Signature in the Early Universe via Ground-Based Laser Interferometers}",
  journal = {Phys. Rev. Lett.},
archivePrefix = "arXiv",
   eprint = {0707.0535},
     year = 2007,
    month = Sep,
   volume = 99,
   number = 12,
      eid = {121101},
    pages = {121101},
      doi = {10.1103/PhysRevLett.99.121101},
   adsurl = {https://ui.adsabs.harvard.edu/abs/2007PhRvL..99l1101S}
}

@article{eastherlim,
   author = {R. {Easther} and E.~A. {Lim}},
    title = "{Stochastic gravitational wave production after inflation}",
  journal = {JCAP},
   eprint = {astro-ph/0601617},
     year = 2006,
    month = Apr,
   volume = {2006},
   number = {4},
      eid = {010},
    pages = {010},
      doi = {10.1088/1475-7516/2006/04/010},
   adsurl = {https://ui.adsabs.harvard.edu/abs/2006JCAP...04..010E}
}

@article{caldwellallen,
    author = "R. R. Caldwell and B. Allen",
    title = "{Cosmological constraints on cosmic string gravitational radiation}",
    reportNumber = "WISC-MILW-91-TH-14",
    doi = "10.1103/PhysRevD.45.3447",
    journal = "Phys. Rev. D",
    volume = "45",
    number = {10},
    pages = {3447--3468},
    year = "1992",
    month = may
}

@article{DV1,
   author = {T. {Damour} and A. {Vilenkin}},
    title = "{Gravitational Wave Bursts from Cosmic Strings}",
  journal = {Phys. Rev. Lett.},
   eprint = {gr-qc/0004075},
     year = 2000,
    month = Oct,
   volume = 85,
   number = {18},
    pages = {3761--3764},
      doi = {10.1103/PhysRevLett.85.3761},
   adsurl = {https://ui.adsabs.harvard.edu/abs/2000PhRvL..85.3761D}
}

@article{DV2,
   author = {T. {Damour} and A. {Vilenkin}},
    title = "{Gravitational radiation from cosmic (super)strings: Bursts, stochastic background, and observational windows}",
  journal = {Phys. Rev. D},
   eprint = {hep-th/0410222},
     year = 2005,
    month = Mar,
   volume = 71,
   number = 6,
      eid = {063510},
    pages = {063510},
      doi = {10.1103/PhysRevD.71.063510},
   adsurl = {https://ui.adsabs.harvard.edu/abs/2005PhRvD..71f3510D}
}

@article{cosmstrpaper,
   author = {X. {Siemens} and V. {Mandic} and J. {Creighton}},
    title = "{Gravitational-Wave Stochastic Background from Cosmic Strings}",
  journal = {Phys. Rev. Lett.},
   eprint = {astro-ph/0610920},
     year = 2007,
    month = Mar,
   volume = 98,
   number = 11,
      eid = {111101},
    pages = {111101},
      doi = {10.1103/PhysRevLett.98.111101},
   url = {https://ui.adsabs.harvard.edu/abs/2007PhRvL..98k1101S}
}

@article{PhysRevLett.106.241101,
  title = {Inspiral-Merger-Ringdown Waveforms for Black-Hole Binaries with Nonprecessing Spins},
  author = {P. Ajith and M. Hannam and S. Husa et al.},
  journal = {Phys. Rev. Lett.},
  volume = {106},
  number = {24},
  pages = {241101},
  numpages = {4},
  year = {2011},
  month = Jun,
  publisher = {American Physical Society},
  doi = {10.1103/PhysRevLett.106.241101},
  url = {https://link.aps.org/doi/10.1103/PhysRevLett.106.241101}
}

@article{Ferraiuolo2025,
	author = {{S. Ferraiuolo} and {S. Mastrogiovanni} and {S. Escoffier} et al},
	title = {Inferring astrophysics and cosmology with individual compact binary coalescences and their gravitational-wave stochastic background},
	DOI= "10.1051/0004-6361/202555124",
	url= "https://doi.org/10.1051/0004-6361/202555124",
	journal = {A\&A},
	year = 2025,
    month = sep,
	volume = 701,
	pages = "A36",
    archivePrefix = {arXiv},
    eprint = {2503.14686},
    primaryClass = {astro-ph.CO}
}
\end{document}